\documentclass[twocolumn]{aastex701}
\usepackage{amssymb}
\usepackage{amsmath}
\usepackage{caption}
\usepackage{subcaption}
\usepackage{color}
\usepackage{lineno}
\linenumbers
\revised{Feb.10.2026}

\submitjournal{ApJ}
\journalinfo{}
\begin{document}

\title{Odd Radio Circles Modeled by Shock-Bubble Interactions}
\author[0000-0002-3143-4552]{Yiting Wang}
\affiliation{Department of Physics, University of Wisconsin-Madison, 1150 University Avenue, Madison, WI 53706, USA}
\email{wang2632@wisc.edu}

\author[0000-0002-8433-8652]{Sebastian Heinz}
\affiliation{Department of Physics, University of Wisconsin-Madison, 1150 University Avenue, Madison, WI 53706, USA}
\affiliation{Department of Astronomy, University of Wisconsin-Madison, 475 N Charter St, Madison, WI 53726, USA}
\email{sheinz@wisc.edu}

\begin{abstract} 
The physical nature and origins of the newly discovered class of Odd Radio Circles (ORCs) remain unclear. We investigate a model whereby ORCs are synchrotron-emitting vortex rings formed by the Richtmyer-Meshkov instability (RMI) when a shock interacts with a low-density fossil radio lobe in the intergalactic medium using 3D magnetohydrodynamic simulations. These rings initially exhibit oscillatory behavior that damps over time. We implement a new method to model Inverse-Compton cooling and synchrotron cooling at high frequencies in a scale-free manner, enabling us to test a wide range of model parameters against the observational constraints. We find that shock strengths of Mach 2-4 are consistent with the data, as expected in accretion, merger-driven, or active galactic nuclei-driven shocks. We find that the initial size of the bubbles required to explain the rings ranges from 140 to 250 kpc, with initial energy in the bubble of order $10^{57}-10^{59}$ erg, consistent with fossil lobes inflated by moderately powerful radio galaxies. Derived ambient pressures and densities place ORCs in low density environments, such as the outskirts of galaxy groups with ages of order 70–200 Myr. Our synthetic radio maps match the polarization properties of ORC1 and predict a dependency of the tangential magnetic field angle on the aspect ratio of ORCs. A key distinguishing trait of the RMI-driven vortex ring model is that it does not require the ORC to be centered on its host galaxy and is therefore redshift agnostic.
\end{abstract}

\keywords{galaxies: clusters: general – radio continuum: galaxies - magneto-hydrodynamics – ISM: bubbles – methods: numerical shock waves}

\section{Introduction} \label{sec:intro}
\subsection{Odd Radio Circles}

Odd Radio Circles (ORCs) are a newly discovered class of ring-like radio sources. The first three ORCs (ORC 1, 2, 3) were observed at 944MHz in 2020 in the Evolutionary Map of the Universe (EMU) Pilot Survey by the Australian Square Kilometre Array Pathfinder (ASKAP) telescope as well as Australia Telescope Compact Array and Murchison Widefield Array \citep{norris_unexpected_2021}.
The fourth ORC (ORC 4) was later found in the data taken by the Giant MetreWave Radio Telescope in 2013\citep{norris_unexpected_2021} at 325MHz. The fifth ORC (ORC 5) was found in ASKAP later by \citep{koribalski_discovery_2021} at 944MHz. Higher resolution data of ORC 1 were given by MeerKAT radio images in \citep{norris_meerkat_2022}. 

They all have comparable diameters of around $1$ arcmin, and the integrated flux density of order several mJy at 944MHz. The feature steep spectral indices of about $-0.5$ to $-1.7$. They are located at high Galactic latitudes of about $\pm 40^\circ$ with redshifts of the presumed central galaxies of $0.3$ to $0.5$.  

In 2023, \citep{koribalski_meerkat_2024} reported the discovery of another new ORC, ORC 6, within the same MeerKAT data. Since then, people have been showing great interest in the observations of ORCs, leading to the identification of other ring-like radio sources as potential ORC candidates. For instance, the SAURON survey \citep[][SAURON]{lochner_unique_2023} and the Coverleaf source\citep[][Cloverleaf]{bulbul_galaxy_2024} both detected objects with similar ring-like morphology, albeit with distinct structures. In 2024, \citep{norris_meerkat_2024} MIGHTEE found another ORC, which is 2-3 times smaller in size and much fainter than the previous ORCs.

Although the radio emission is known to be synchrotron emission, the source of the plasma and its relation to the nearby galaxies remain unclear, leading to a wide range of proposed formation scenarios. \citet{norris_meerkat_2022} considered the ORCs as a spherical shell of synchrotron emission from the starburst termination shock in the host galaxy. \citet{dolag_insights_2023} simulated the merger shocks from the galactic halos and found that the massive halos present similar features as the ORCs. Another report \citep{coil_ionized_2024} shows that spectroscopic data of ORC4 exhibits strong [O II] emission which extends over 40kpc. The velocity gradient and high-velocity dispersion across the [O II] nebula suggest that the shock-ionized gas is moving toward the galaxy due to a shock caused by a strong outward-moving wind from the galaxy. They argue that the radio emission in ORCs may result from the same starburst wind. A recent paper \citep{lin_active_2024} investigated the possibility of the ORC as end-on views of cosmic-ray proton-dominated bubbles.

\subsection{The Richtmyer-Meshkov Instability as the ORC Origin}
\citet{shabala_are_2024} proposed that ORCs may be formed by shock-bubble interactions when radio lobes interact with intergalactic shock waves. Their work builds on the idea that radio relics may be formed as the results of such interactions \citep{ensslin_formation_2002,heinz_heating_2005,friedman_all_2012,Nolting:2019lsk}. The theoretical foundation for these interactions originates from the Richtmyer-Meshkov instability (RMI), first described by \citet{richtmyer_taylor_1960} and later experimentally validated by \citet{meshkov_instability_1969}. 

In this paper, we adopt the idea that ORCs result from synchrotron-emitting vortex rings produced by shock-bubble interactions in the intergalactic medium (IGM) as our working hypothesis. We conduct a detailed parameter study of this process to constrain the possible environments and energetics and to derive testable predictions that can differentiate this process from other proposed ideas for the origins of ORCs. 

The Richtmyer-Meshkov instability (RMI) was first described by \citet{richtmyer_taylor_1960} and later experimentally validated by \citet{meshkov_instability_1969}. Richtmyer’s analysis highlighted the growth of perturbations at the interface between two fluids of different densities when impulsively accelerated by a shock wave, leading to the characteristic instability known as RMI. Meshkov extended this work by experimentally confirming the instability's behavior when a shock wave traverses the interface between two gases of differing densities, demonstrating that the interface disturbance grows linearly with time. We note that the RMI is often described as the impulsive limit of the Rayleigh-Taylor instability, and thus the formation of the vortex ring is functionally analogous to the formation of vortex rings in buoyantly rising bubbles.

Building on this foundational work, \citet{quirk_dynamics_1996} conducted a detailed numerical study on the dynamics of shock-cylindrical bubble interactions, revealing how shock waves compress and distort bubbles, generating persistent vortex rings. \citet{ranjan_experimental_2005} further explored this phenomenon through experimental investigations, where a strongly shocked spherical gas bubble subjected to a Mach 2.88 shock in atmospheric nitrogen was studied. Their experiments showed that vorticity deposited on the bubble surface during shock interaction leads to the formation of major and secondary vortex rings, visualized for the first time using laser sheet imaging. They further advanced the understanding of these interactions in \citet{ranjan_shock-bubble_2011} where they highlighted the formation of vortex pairs in 2D and vortex rings in 3D cases, showing that these structures are critical to the dynamics of shock-bubble interactions.

\citet{niederhaus_computational_2008} expanded the investigation by conducting a computational parameter study for three-dimensional shock-bubble interactions. Their work emphasized the role of different Mach numbers, bubble sizes, and ambient conditions in determining the morphology and stability of the resulting vortex rings. For higher Mach numbers, the primary vortex core becomes indistinguishable.

The application of the shock-bubble interaction concept in astrophysical contexts was first explored by \citet{ensslin_formation_2002} in their study of cluster radio relics. They investigated how radio plasma cocoons, left over from past active galactic nucleus (AGN) activity, could be re-energized by merger shock waves passing through galaxy clusters. Their 3D magnetohydrodynamic (MHD) simulations demonstrated that when these radio cocoons encounter shock waves, they are compressed and can be transformed into filamentary or toroidal structures, similar to those observed in radio relics. This process not only revives the radio emission but also enhances the polarization of the radio waves, producing a characteristic morphology that aligns well with observations of relics in galaxy clusters.

Building on this foundational work, \citet{heinz_heating_2005} conducted hydrodynamic simulations to explore the role of weak shocks and sound waves in heating the intracluster medium (ICM) through interaction with bubbles of relativistic gas. Their study revealed that these bubbles, when impacted by shocks or sound waves, act as catalysts for transforming kinetic energy into heat via the formation of vortex rings and other turbulent structures. The study confirmed the efficiency of vortex formation due to the Richtmyer-Meshkov instability (RMI) and highlighted how the kinetic energy generated by these vortices could contribute significantly to the heating of the ICM, thereby maintaining the observed thermal structure of galaxy clusters without the need for continuous energy input from external sources.

In this paper, we present a comprehensive study of radio ring emission modeled by shock-bubble interactions in 3D magnetohydrodynamic (MHD) simulations, proposing this mechanism as a potential origin for the observed ORCs, complementing the work by \citet{shabala_are_2024}. By design, as the shock hits the bubble, it initiates a compression of the bubble; due to the higher sound speed inside the bubble, the compression travels more rapidly through the bubble, which thus evolves into a differentially rotating vortex through the RMI\citep{richtmyer_taylor_1960,meshkov_instability_1969,heinz_heating_2005,friedman_all_2012}. 

Section \ref{sec:method} details the simulation setup, initial conditions, and analysis techniques employed in our study. We describe the parameters chosen for the simulations, such as the magnetic field configurations, shock velocities, and the properties of the ambient medium. Section \ref{sec:analysis} presents the results of our simulations, providing a range of visualizations and quantitative analysis. We highlight the formation and evolution of the radio rings, showcasing the morphological and kinematic properties of the emission. In Section \ref{sec:discussion}, we interpret our results and draw connections between the simulated radio rings and the observed ORCs. We discuss the implications of the simulations on the physical radius, flux density, and the pressure environment of the ORCs. Finally, Section \ref{sec:conclusion} provides a concise summary of our study.

Throughout this paper, we assume standard $\Lambda$CDM condcordance cosmology with parameters chose from \citet{planck_2013}.

\section{Methods} \label{sec:method}

\subsection{Simulation Setup}
The goal of the simulations presented in this paper is to test the hypothesis that some or all ORCs might be vortex rings generated by shock-bubble interaction, as also proposed in \citet{shabala_are_2024}. To this end, we conduct simulations of shock-bubble interactions using {\texttt{FLASH}} 4.6.2. The simulations are run in 3D Cartesian coordinates. We solve the magnetohydrodynamic (MHD) equations using the Unsplit Staggered Mesh (USM) solver and the Constrained Transport (CT) method  implemented in \texttt{FLASH}, which ensures that the simulation maintains the divergence-free condition of the magnetic field to machine precision. 

The initial conditions are defined by a magnetized spherical bubble that is subject to a planar shock. The shock travels along the z-direction toward the bubble along a grid with dimensions of 20x20x200 bubble radii. 

Our base grid comprises one 8x8x8 computational block in the $x$-direction, one block in the $y$-direction, and ten blocks in the $z$-direction. We use adaptive mesh refinement (AMR), permitting the grid resolution to adapt dynamically in regions of interest. For most of our simulations, the maximum allowable refinement level is 8, which corresponds to a resolution of 128 cells across the bubble diameter. The time-dependent refinement criteria follow the standard second-derivative based FLASH scheme, using the fluid pressure (to resolve the shock) and the tracer fluid of the bubble (to ensure the fluid interface between the bubble/vortex and the environment is resolved).

The simulations have reflecting boundary conditions in the $x$ and $y$ directions, while they have inflow-outflow boundary conditions in the z-direction.

The Atwood number (the ratio of environment density to bubble density) is set to $A\equiv\rho_{\rm IGM}/\rho_{\rm bubble}=100$\footnote{Note that the dynamics of shock-bubble interactions are insensitive to $A$ as long as $A\gg 1$.}, making the bubble underdense, corresponding to the underlying model of a fossil radio lobe or similar bubble being overrun by a shock.

For the most part, our simulations are scale-free, however, the synchrotron and inverse Compton cooling applied to our tracer particles used to calculate emission properties (described below) have different dependencies on the scale-setting simulation parameters (bubble radius, sound speed, and pre-shock density). Therefore, some care is warranted when considering synchrotron spectra at higher frequencies, as discussed in \S\ref{sec:synchrotron}.



The simulations employ a multi-gamma adiabatic equation of state with an adiabatic index (ratio of specific heat) of $\gamma_{\rm ad}=4/3$ inside the bubble, assuming a relativistic non-thermal fossil plasma, and $\gamma_{\rm ad}=5/3$ for the background gas. We employ two passively advected tracer fluids for the environment and the bubble fluid.

We impose a shock moving in $z$ direction. We investigate the impact of a range of Mach numbers, including $1.4$, $2$, $4$, $8$, and $16$, representing diverse shock strengths in \S\ref{subsec:differentmachs}.

A resolution study is presented in \S\ref{subsec:differentresols}. We further investigate the effect of different magnetic topologies \S\ref{subsec:differentcuts} on the evolution and observable properties of shock-bubble-generated radio circles in later sections.

\begin{figure*}
\centering
\includegraphics[width=1\linewidth]{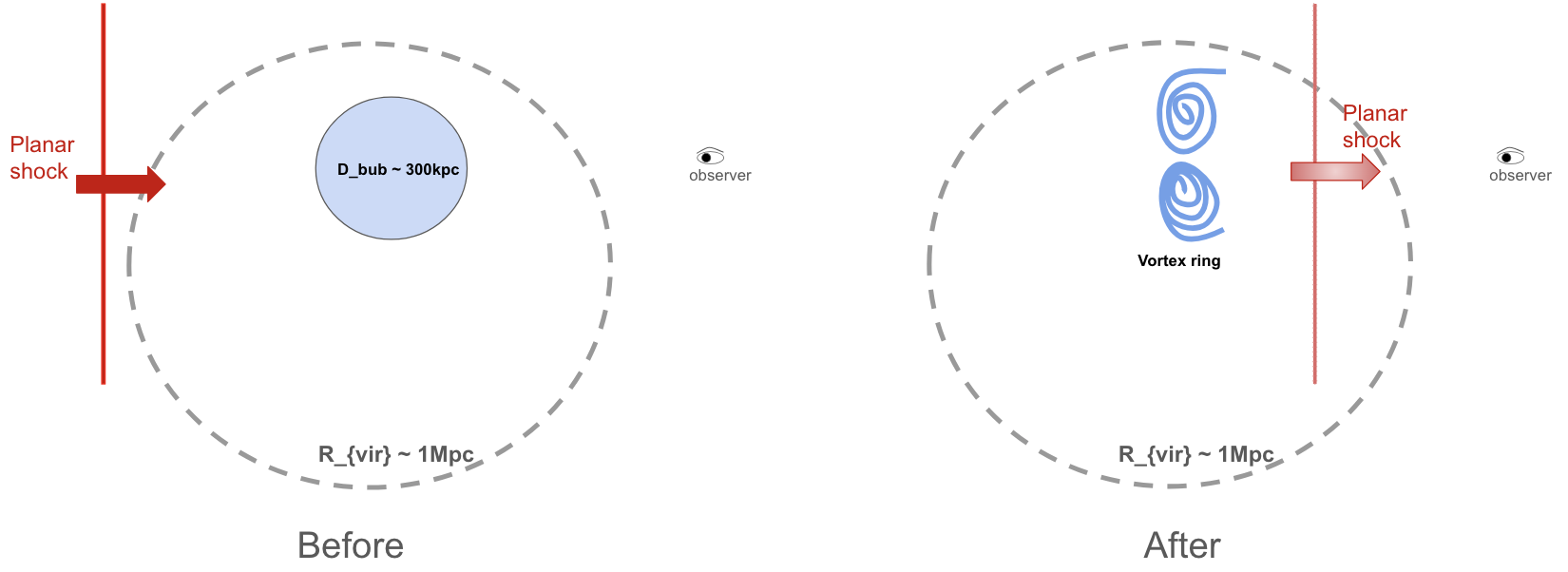}
\caption{Schematic illustration of the shock-bubble interaction setup mapped to a realistic galaxy group environment. (Before) A planar shock propagates through the ambient medium and hits a fossil radio bubble of diameter $D_{\rm bub}\sim300$ kpc embedded within a group halo of virial radius $R_{\rm vir}\sim1$ Mpc. (After) The interaction generates vorticity, producing a vortex-ring-like structure that can give rise to a ring-like radio morphology in projection. The eye symbol indicates the line of sight. The schematic is not to scale except where indicated and is intended to illustrate geometry rather than detailed dynamics.}
\label{fig:shock_bubble_schematic}
\end{figure*}

Since, observationally, we are only interested in the radio ring emission, we set the external magnetic field strength to zero, while the initial average plasma beta (the ratio of thermal pressure to magnetic pressure) inside the bubble is set to $\beta=10$. To sample different possible field topologies inside the bubble, we loosely followed the method described by \citet{ruszkowski_impact_2007}. Specifically, we generated stochastic magnetic fields by setting up each component of the magnetic vector potential in Fourier space to satisfy a given power spectrum and assign a uniformly distributed random phase to each Fourier component, such that the inverse Fourier transform generates a randomized vector potential with the desired statistical properties. We impose a boundary condition on the vector potential to ensure that the magnetic field is tangential at the bubble surface and set the field to zero outside of the bubble. We then calculate the fully divergence-free magnetic field from the vector potential. We choose the power spectrum for $\vec{A}$ such that the field has a characteristic coherence length $\lambda_{\rm max}=2\pi/k_{\rm min}$ and that the power spectrum falls with a set powerlaw index of $\kappa = -2.5$ to an upper cutoff at $k_{\rm max}$.

Table 1 presents a summary of the simulation parameters we varied and the physical parameters we investigated in each case. The lower left part of the table shows the initial simulation parameters that are kept constant to limit the scope of this research. In \S\ref{sec:analysis}, we present the results from our grid of simulations.

\begin{table}
    \centering
    \begin{tabular}{l|l}
      \textbf{Simulation Parameter}      & \textbf{Physical Outcome} \\
      \hline
      Shock Mach number  & Ring radius, width \\
      \ \ \ \ \  (1.4, 2, {\color{gray}{2.83}}, {\color{blue}{4}}, 6, 8, 16)  \\
      Resolution (64, {\color{blue}{128}}, 256)             & Ring width convergence \\
      Field coherence length                 & Polarization fraction \\
      (1/8, 1/4, 7/20, {\color{blue}{1/2}}, 7/10, 1 D$_b$)                         & \\
      Field seed ({\color{blue}{1}}, 2, 3)                  & Magnetic field angle \\
      Box/bubble radius (10)                 & Flux density \\
      Atwood number (0.98)                  & Magnetic field strength \\
      Bubble $\gamma_{\rm ad}$ (1.33)                & Adiabaric index \\
      IGM $\gamma_{\rm ad}$ (1.67)                 & \\
      Bubble plasma $\beta$ (10)                   & \\
      \hline
    \end{tabular}
    \caption{Summary of the main simulation parameters (left column) and physical quantities analyzed (right column) in this study affected by the relevant parameters. Parameter choices for our fiducial run are listed in blue. The Mach 2.83 simulation was used only in the model fitting described in \S\ref{sec:discussion_modelfitting}. Resolution refers to the number of computational cells per bubble diameter, field seed refers to three simulations with different seeds used to initialize the B-field but otherwise similar parameters as our fiducial run.}
\end{table}

\subsection{Synchrotron emission}

To model the radiative properties of the vortex rings generated in our simulations, we follow the method outlined in \citet{chen_numerical_2023} \citep[see also][]{vaidya:18}. We employ Lagrangian tracer particles inside the bubble to track radiative and adiabatic cooling of the non-thermal fluid. Each tracer particle represents the local distribution of non-thermal particles and is used to solve the particle transport equation  implicitly along its trajectory through the fluid {\citep[e.g.,][]{kardashev:62,bicknell:82,coleman:88,vaidya:18,chen_numerical_2023}}. 

This solution includes adiabatic changes and synchrotron and inverse Compton cooling off the cosmic microwave background (synchrotron-self-Compton losses and losses off the intergalactic background light can be neglected for diffuse plasmas like the ones simulated here and assumed to be responsible for the radio emission from ORCs.)

We then calculate the synchrotron emission using the tracer particles to incorporate radiative cooling. Throughout, we assume a particle energy spectral index of electron distribution of $s=2$ such that the electron distribution satisfies $f(\gamma) =f_{0}(\gamma/\gamma_{0})^{-s}$. 

The low-frequency emission is unaffected by cooling and can be normalized entirely from the fluid grid properties \cite[non-thermal pressure and the perpendicular magnetic field strength, following, e.g., ][]{chen_numerical_2023}. 

We employ nearest-neighbor matching to calculate the cooled synchrotron spectrum and calculate Stokes I, Q, and U fields by ray-tracing the synchrotron emissivity through the simulation cube to generate total intensity and polarization images.

\subsubsection{Radiative Cooling and Spectral Calculation}
\label{sec:synchrotron}
In general, the synchrotron spectrum of a cooled electron distribution can be calculated from the solution of the transport equation, using the method of characteristics described above. The solution can then be cast in terms of the injection spectrum (assumed to be a powerlaw) and the characteristic mapping of equation 14 from \citet{chen_numerical_2023}:
\begin{equation}
    \gamma_{\rm c}\left(\gamma_{0}.t\right)=\frac{\left(\rho(t)/\rho_0\right)^{1/3}}{1/\gamma_{0} + \int_{t_{0}}^{t}dt' A(t')\left[\rho(t')/\rho_{0}\right]^{1/3}}
    \label{eq:cool}
\end{equation}
where $A=\frac{4\sigma_{\rm T}}{3m_{\rm e}c}\left(U_{\rm B} + U_{\rm CMB}\right)$. The characteristic cooling energy is then set by taking the limit $\gamma_{0}\rightarrow \infty$. In the spectrum, this characteristic energy is expressed as the characteristic synchrotron frequency
\begin{equation}
    \nu_{\rm c}=\frac{\gamma_{\rm c}^2 e B}{m_{\rm e} c}
    \label{eq:nuc}
\end{equation}

Because we aim to use our simulations in a scale-free manner, we cannot simply scale the synchrotron spectrum as we change the physical parameters, as it can be see from eq.~\ref{eq:cool} that inverse Compton cooling and synchrotron cooling depend on the scale parameters in different ways, while adiabatic cooling is scale invariant. In order to still allow us to choose different scale parameters, we calculate two sets of spectra--one for synchrotron cooling only and one for inverse Compton cooling only. Each of these spectra can then be re-scaled appropriately as follows:

We know that, in the absence of radiative cooling, the spectrum will have a simple powerlaw shape,
\begin{equation}
    F_{\nu}(t=t_0)=F_{\rm norm}\nu^{-\alpha}
    \label{eq:powerlaw}
\end{equation}
with the normalization $F_{\rm norm}$
\begin{equation}
    F_{\rm norm} \propto \frac{P, V, B^{1+\alpha}}{D_{A}^{2},\nu_{0}^{\alpha}}
\end{equation}

For reference, we set $t_{0}$ to be equal to the time the shock encounters the bubble, assuming that the electrons responsible for the observed emission are accelerated or re-energized by the interaction with the shock through, e.g., first order Fermi acceleration.

We also know that, in the presence of radiative cooling, the low frequency spectrum will still follow the same power law, while the integrated spectrum will be changed by some characteristic dimensionless shape function $f(\nu/\nu_{\rm c})$ such that
\begin{equation}
    F_{\nu}=F_{\rm norm}\cdot f\left(\frac{\nu}{\nu_{\rm c}}\right)\cdot\nu^{-\alpha}
\end{equation}
Functionally, we can evaluate $f$ by dividing the calculated spectrum for a given set of scale parameters by the powerlaw from eq.~(\ref{eq:powerlaw}) and use the known scaling of $\nu_{\rm c}$ from eqs.~(\ref{eq:nuc}) and (\ref{eq:cool}) either for synchrotron cooling or inverse Compton cooling.

We note that this is an approximation of the true spectrum. However, given that it allows us to cover a very large parameter space that our simulations are applicable to, we use it with the caveat that the resulting observational constraints carry some uncertainty. Given the overall uncertainties carried by, e.g., source distance, we posit that this is acceptable compromise.

\subsubsection{Polarization}

The polarization fraction $\Pi$ is defined as the ratio of the polarized intensity to the total intensity (Stokes parameter I), where Q and U are the Stokes parameters representing the linear polarization components. 
\begin{equation}
 \Pi_{pol}=\frac{\sqrt{Q^2 + U^2}}{I} 
\end{equation}
The polarization angle is defined using Stokes parameters
\begin{equation}
    \theta_{pol} = \arctan{\left ( \frac{U}{Q+\sqrt{U^2+Q^2}}\right )} 
\end{equation}
The B-field angle is rotated by 90$^{\circ}$ relative to the polarization vector. 

Given the axial ring-line morphology of ORCs and the vortex rings simulation here, we measure the polarization angle {\rm relative to the radius vector} from the center of the ring 
\begin{equation}
    \phi_{pol} = \arccos(\left| v_x r_x+v_y r_y \right|) 
\end{equation}
and therefore, magnetic field angles are measured relative to the tangent vector of the ring.

We calculate emission-weighted polarization angles as
\begin{equation}
    \langle \psi_{B} \rangle = \frac{\sum (\phi_B \cdot I)}{\sum I} \times \frac{180}{\pi}
\end{equation}

\subsection{Applying a choice of scale parameters}

In order to allow direct comparison with observations, we must choose a set of scale parameters. This is most transparently done by choosing a physical length scale for the radius of the bubble, $R_{0}$, a physical value of the adiabatic sound speed $c_{s}$, and either a density or pressure scale for the initial circum-bubble medium (corresponding to the pressure or density of the environment), with the following relations for dependent quantities:
\begin{align}
    c_s &= c_{s,0} \sqrt{\frac{P}{P_0}} \sqrt{\frac{\rho_0}{\rho}}\\
    t &= t_0 \left(\frac{R}{R_0}\right) \left(\frac{c_{s,0}}{c_s}\right)\\
    B &= B_0 \sqrt{\frac{P}{P_0}}\\
    \gamma_{\rm c}^{\rm CMB}&= \gamma_{\rm c,0}^{\rm CMB} \frac{t_0}{t} \\
    \gamma_{\rm c}^{\rm synch}&= \gamma_{\rm c,0}^{\rm synch} \frac{t_0}{t} \frac{P}{P_0}\\
    \nu_{\rm c}&= \nu_{\rm c,0} \frac{B}{B_0}\left(\frac{\gamma_{\rm c}}{\gamma_{\rm c,0}}\right)^2\\
    F_\nu &= F_{\nu,\rm norm} \cdot f\left(\frac{\nu}{\nu_c}\right) \cdot \nu^{-\alpha}\\
    F_{\nu,\rm norm} &= F_{\nu,\rm norm,0} \left(\frac{R}{R_0}\right)^3 \left( \frac{B}{B_0}\right)^{3.5} \left(\frac{\nu}{\nu_0} \right)^{-\alpha}
\end{align}
%
%
%
%
where the subscript 0 denotes reference simulation values (chosen to calculate the nominal cooling parameters in the simulations).

Our spectral calculations cover a frequency range of $\nu_{\rm max}/\nu_{\rm min} = 2.7\times 10^{6}$ in order to facilitate rescaling our simulations over a wide range of parameter choices and still capture spectral evolution at the observational frequency for the ORCs analyzed in this paper.

For any given set of scale parameters, we then calculate the flux density as a function of frequency, properly transformed to the assumed source redshift, and interpolate to the observing frequency on a log-log grid of flux and frequency.


\section{Results}\label{sec:analysis}

Our 3D MHD simulations reveal a set of robust patterns in the formation and evolution of vortex rings resulting from shock-bubble interactions. Across a range of shock Mach numbers, our simulations demonstrate the creation of vortex rings that persist over many shock crossing times. This aligns with previous studies on shock-bubble dynamics \citep[e.g.][]{ranjan_shock-bubble_2011,niederhaus_computational_2008}.

For reference, we choose the Mach 4 simulation as our fiducial case for a detailed comparison. The Mach 4 case provides a clear and representative example of the shock-bubble interaction process for an astrophysically realistic strong shock. This particular run showcases the characteristic ring-like morphology that is typical of the vortex rings formed in our simulations. By using this fiducial simulation as a reference point, we can effectively highlight and discuss the variations observed in other runs with different shock Mach numbers in the later sections. Throughout this paper, we refer back to the Mach 4 case to illustrate standard behaviors and structural features, using it as a baseline to point out and analyze differences and deviations in the other simulations, which allows us to systematically explore the parameter space and better understand the influence of varying initial conditions on the resulting physical properties of the vortex rings.

\subsection{Fiducial Case: Mach 4 Simulation}\label{subsec:Mach4}

Our fiducial simulation features a shock Mach number of 4, an average initial plasma beta of  $\langle\beta\rangle=10$, and short wavelength cutoffs of a quarter bubble diameter and a long-wavelength cutoff (coherence length) of one half bubble diameter for the magnetic power spectrum. That is, the coherence length of the magnetic field is smaller than, but comparable to the bubble size. We vary the magnetic power spectrum to investigate its impact on the observational properties of the simulated ORCs.

Simulations at different shock strengths show qualitatively similar behavior, with quantitative differences we will discuss below.


The initial analysis (used to describe the overall properties of the vortex) focuses on the lowest frequency in our spectral grid, well below any effects of radiative cooling (verified by inspection of the spectral index, which equals the initial injection value of $\alpha=0.5$.) This allows us to emphasize the resulting patterns of the simulations, which remain consistent across different frequencies when cooling is not considered. It is crucial to note that our observational comparisons are made at higher frequencies, corresponding to the actual observing frequencies of the different ORCs listed in Tab.~\ref{tab:ORCs}.

\subsubsection{Total Intensity}
\begin{figure*}
    \centering
    \includegraphics[width=\linewidth]{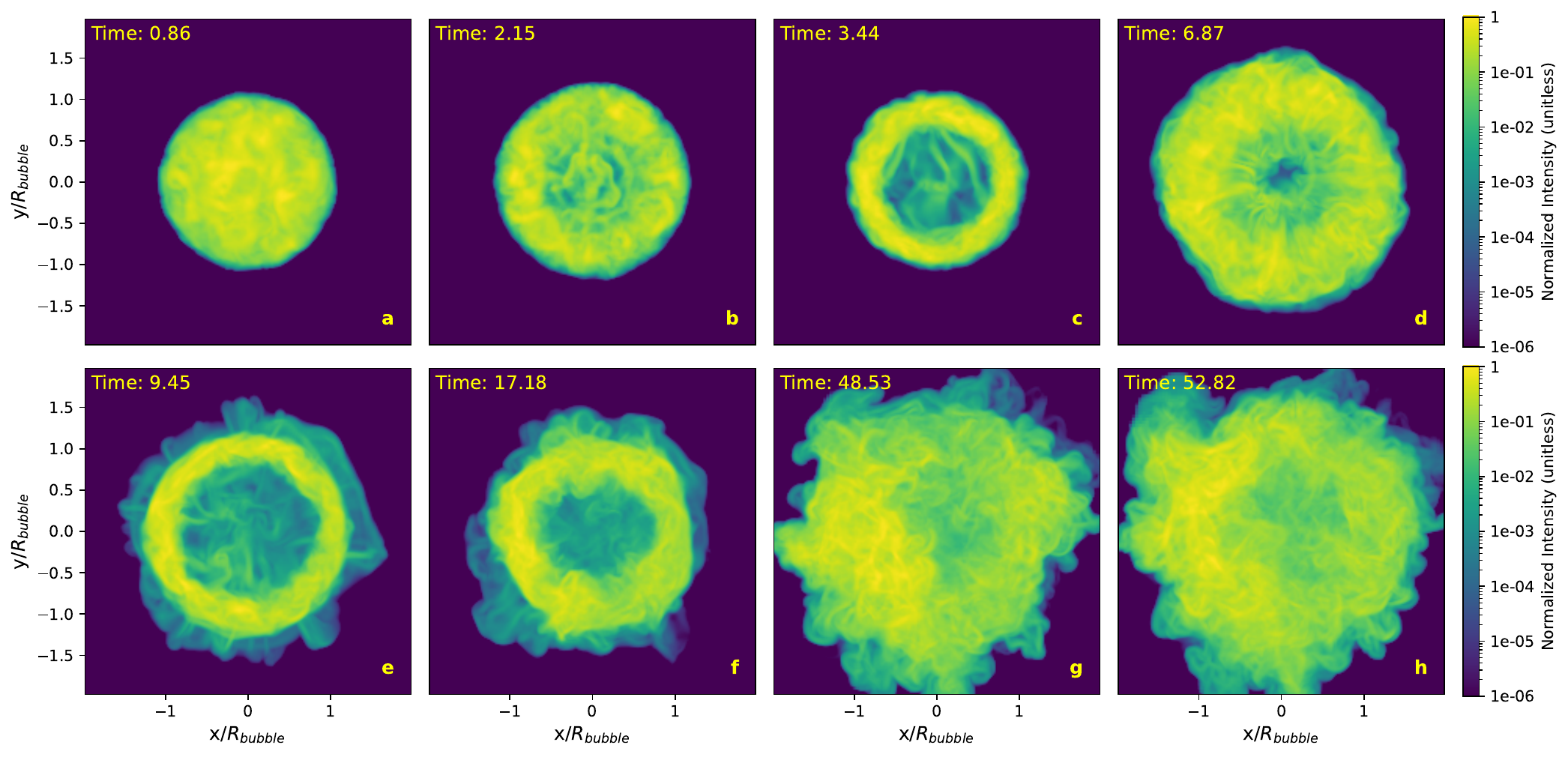}
    \caption{
    Synthetic low-frequency synchrotron intensity (Stokes I) for the fiducial Mach 4 vortex ring simulation, shown face-on at eight evolutionary stages (a-h). 
    \textbf{Color scale}: Logarithmic, unitless normalized intensity (with each panel scaled $I/I_{\text{max}}$), using the \texttt{viridis} colormap. Both rows share the same normalized intensity range ($10^{-6}$ to $1$). 
    \textbf{Spatial coordinates}: Axes in units of bubble radius ($x/R_{\rm bubble}$, $y/R_{\rm bubble}$), spanning $\pm2R_{\rm bubble}$. The central bubble is positioned at (0,0). 
    \textbf{Temporal evolution}: Times in shock crossing units ($t_{\rm cross} = R_{\rm bubble}/v_{\rm s}$):
    (a) $0.86 t_{\rm cross}$. 
    (b) $2.15 t_{\rm cross}$, 
    (c) $3.44 t_{\rm cross}$,
    (d) $6.88 t_{\rm cross}$,
    (e) $9.45 t_{\rm cross}$,
    (f) $17.18 t_{\rm cross}$,
    (g) $48.53 t_{\rm cross}$,
    (h) $52.82 t_{\rm cross}$.}
    \label{fig:Intensity}
\end{figure*}

Figure \ref{fig:Intensity} shows face-on synthetic radio images of the bubble/ring at different times after it is hit by the shock wave (at time $t=0$). The angular scales are normalized by the initial bubble radius $R_{bubble}$. While the first inset (during shock passage) does not show any ring-like morphology, all following images show a central intensity depression, i.e., a ring. The images also indicate that the aspect ratio of the ring (the ratio of ring radius to ring width) changes, with maxima (i.e., thinnest ring) in the third and fifth image. This oscillation in the aspect ratio is also reflected in other properties of the ring, as discussed below. We will refer to this initial oscillatory behavior as "breathing".

The figure further shows that the ring is well defined for about 20 shock crossing times $t_{\rm cross}\equiv R_{\rm b}/v_{\rm shock}$, after which it
becomes less well defined and the appearance becomes more and more turbulent, and the width of the ring increases substantially until a ring is no longer discernible. 

We find that the intensity peaks during times when the ring is narrowest.  A quantitative comparison with the observations will follow in \S\ref{sec:discussion}.

The appearance of a circular radio ring happens only for a limited range of viewing angles. By rotating the simulation ring about an axis and looking at the morphology, we find the ring-like structures are recognizable for inclinations within 56.5$^{\circ}$. Integrating over solid angle, this corresponds to a viewing probability of about 45\% of all possible viewing directions. Figure \ref{fig:Intensity} specifically corresponds to a face-on orientation of the shock surface in the shock-bubble scenario, which is the viewing angle we analyzed in this paper. Such orientation shows clear symmetry and maximizes edge brightness, making them easier to identify as ORCs in current radio surveys. For other inclined viewing angles, the same physical configuration would produce distorted or incomplete ring-like features rather than a clean circular morphology. Such systems are more difficult to distinguish from radio arcs, relics, or filamentary emission and may therefore be missing in existing ORC samples.

\subsubsection{Polarization}
\begin{figure*}
    \centering
    \includegraphics[width=\linewidth]{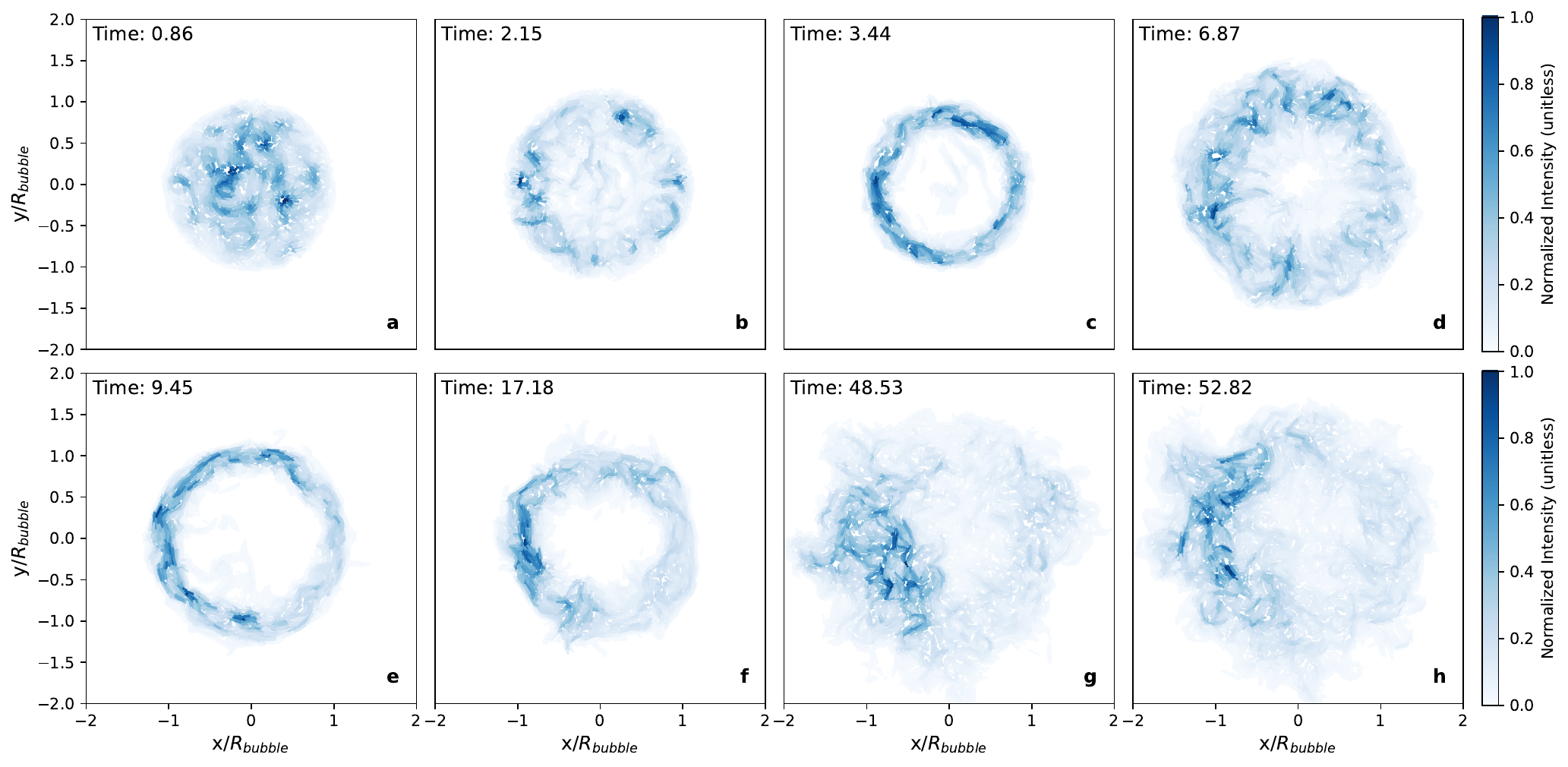}
    \caption{B-field maps of the fiducial Mach 4 simulation at eight evolutionary stages (a-h), corresponding to the intensity sequence in Fig.~\ref{fig:Intensity}. 
    \textbf{Color scale}: Normalized intensity ($I/I_{\text{max}}$) using a 'Blues' colormap (darker shades indicate brighter emission). 
    \textbf{Vectors}: magnetic field orientation (derived by rotating polarization E-vectors by $90^\circ$), with length proportional to polarization fraction $p = \sqrt{Q^2 + U^2}/I$. Vectors are sampled at random pixels where $I > 0.01 I_{\text{max}}$. 
    \label{fig:Polarization}}
\end{figure*}

In addition to the total intensity, Fig.~\ref{fig:Polarization} shows the polarization maps of the fiducial Mach 4 simulation at different times. The footpoints of the polarization quivers are randomly selected from the set of pixels where intensity exceeds 1$\%$ of the maximum intensity pixel at the given time. The color scale represents normalized intensity, while the length and direction indicate polarization fraction and polarization angle, respectively.

The first image provides a good indication of the initial magnetic field configuration inside the bubble. The subsequent images show the on-the-sky magnetic field orientation (derived from rotating the polarization vector by $90^{\circ}$).

The magnetic field quivers show a clear relation to the morphology of the ring. While the polarization vectors carry over some of the stochasticity imprinted by the initial random field topology, the quivers show some alignment with the toroidal direction, which is strongest when the ring is thinnest. At later times, the magnetic vectors appear less ordered, as the ring width increases. Interestingly, during phases when the ring width is large, such as in the fourth frame at time $t=6.67$, the magnetic field displays a substantial radial component.

\subsubsection{Statistical Properties of Observables}

\begin{figure*}
    \centering
    \includegraphics[width=\linewidth]{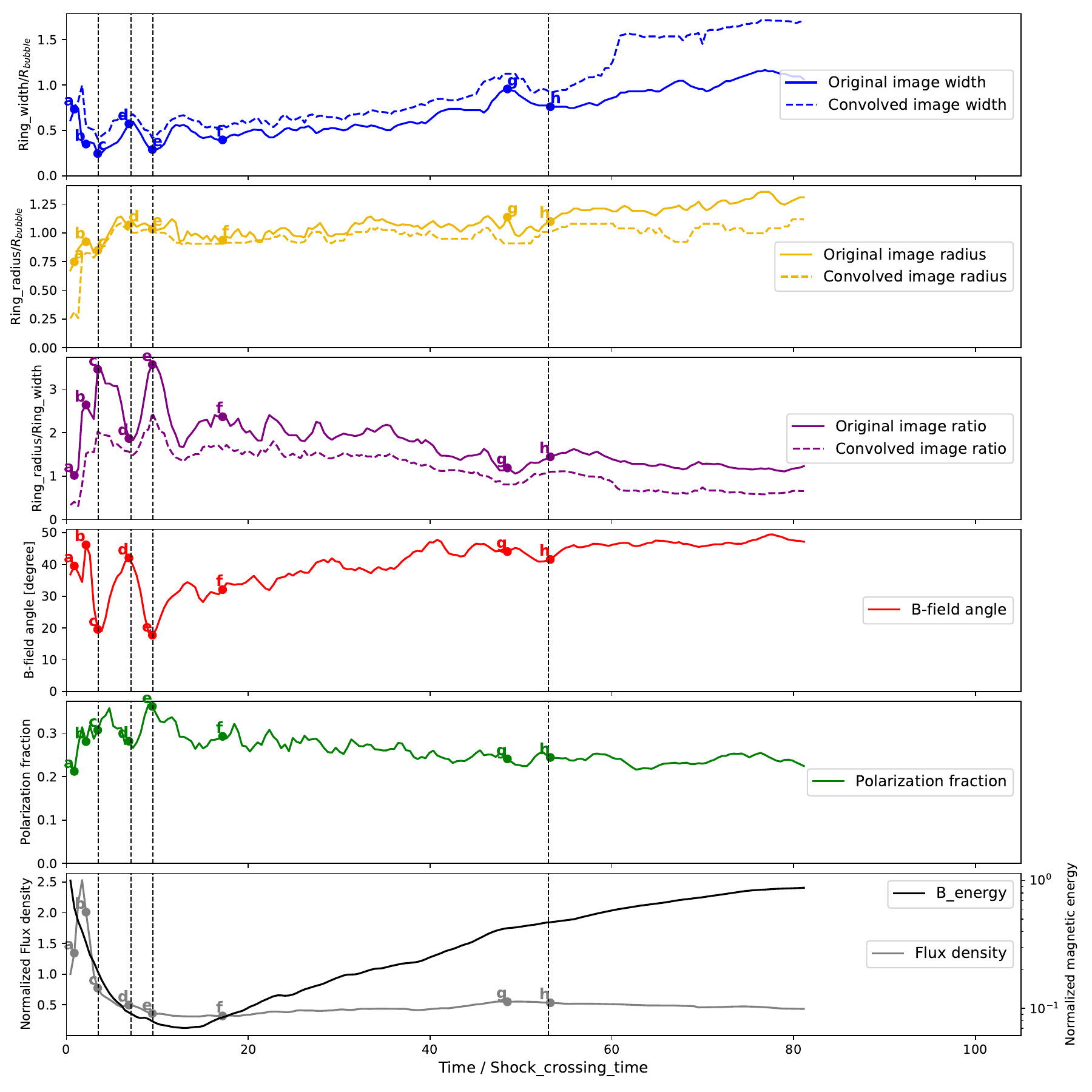}
    \caption{{\em Top to bottom:} Five observables as function of time for our fiducial Mach 4 simulation: ring width, ring radius, (the aspect ratio,) from the original simulation images and the convolved images, polarization fraction, B-field angle, flux density (normalized by initial flux density) and magnetic energy (normalized by the initial magnetic energy) from the original simulation images. The dashed curves show the same diagnostics for images smoothed to the same effective image resolution as ORC5.
    \label{fig:radialplots}}
\end{figure*}

It is clear from the image sequences that the evolution of different observable quantities is related. To evaluate this more quantitatively, and to provide observational diagnostics to compare among different simulations and with current and future ORC observations, we calculate the time evolution of five robust average observable quantities: the ring width, the ring radius, the polarization fraction, the magnetic field angle, and the flux density of the ring, which are shown in Fig.~\ref{fig:radialplots}.  

\begin{figure}
    \centering
    \includegraphics[width=\linewidth]{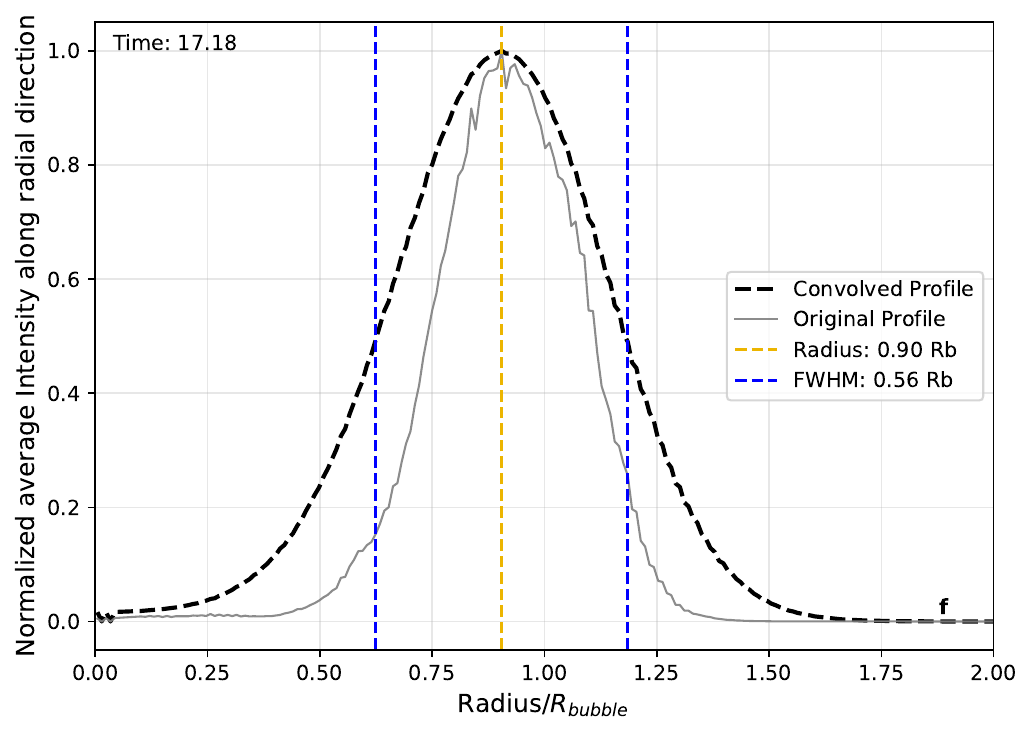}
    \caption{Radial intensity profile (normalized by the peak intensity) at time $t=17.18t_{\rm cross}$ (time stamp {\em  f} in Fig.~\ref{fig:radialplots}.) Shown are the raw intensity profile (dotted blue curve) and the same radial intensity profile of an image smoothed to the same effective resolution as ORC5 (solid blue curve). Also shown are the radius at the peak intensity (dashed green line) and the locations of the FWHM (dashed purple lines).}
    \label{fig:cumulativeintensity}
\end{figure}

To calculate the quantities shown in these plots, we first generate the azimuthally averaged radial radio intensity profiles from the center of the ring in the synthetic radio maps from the original data. Figure \ref{fig:cumulativeintensity} presents the averaged radial intensity profile at $t=17.18$ as an example of our definitions of the observables. The dashed blue lines represent the intensity at each radius bin of the original data, while the solid blue line shows the smoothed intensity after convolution with the telescope beam, representative of the relative resolution of the ASKAP image of ORC5. The dashed green vertical line marks the radius of the ring at peak intensity of the convolved image, and the dashed purple vertical lines indicate the full width at half maximum of the convolved image, defining the width of the ring. In \S\ref{sec:discussion}, we will use the convolved images to discuss the observational predictions and implications of this model.

The panels in Fig.~\ref{fig:radialplots} indicate that the quantities plotted experience initial strong oscillatory behavior that then damps out as the evolution progresses. For example, the ring width shows two clear minima, denoted by the labels {\em c} and {\em e} in the figure and a maximum at time {\em d}. Similarly, the angle of the magnetic field vectors relative to the ring tangent directions shows minima at the times {\rm c} and {\em e}, and a maximum at time {\em d}, while the polarization {\em fraction} shows a minimum at time {\em d} and maxima near times {\em c} and {\em e}. At later times, the amplitude of these oscillations decreases,  but they are still visible in the aspect ratio, shown in the middle panel.

We refer to these correlated initial oscillations of the width, polarization fraction, and the B-field angle in the first 20 shock crossing times as "breathing" for the changing aspect ratio of the rings. This dynamic behavior is also evident in the intensity plots Fig.~\ref{fig:Intensity} and polarization maps Fig.~\ref{fig:Polarization}. 

Concurrently, the overall radius and width of the ring steadily increase. As time progresses, the ring appears to become more turbulent and subject to numerical dissipation as turbulence reaches the resolution scale of our simulations, as can be seen in both the intensity and polarization plots.

The ring width and B-field angle exhibit a strong positive correlation (correlation coefficient $r=0.81$), with closely matched early-time oscillation frequencies of $0.159$ and $0.174$ in units of $t_{\rm cross}^{-1}$, respectively. In contrast, the width and polarization fraction are moderately anti-correlated ($r=-0.67$), and the B-field angle and polarization fraction show a strong negative correlation ($r=-0.77$). This pattern indicates that the polarization fraction oscillates approximately out of phase with the other two quantities, consistent with a phase lag of about $\pi/2$ relative to both the width and B-field angle per ring radius. As the ring width expands and the field reorients, the polarization fraction drops, while the width and field angle increase together. These diagnostics are particularly useful, as all three can in principle be measured directly from radio images of ORCs.

The anti-correlation of the polarization fraction with the other parameters is especially prominent at early times, matching the period of coherent oscillations. At later times, however, the amplitude of oscillations in all three diagnostics diminishes and noise dominates, making it difficult to extract meaningful correlations for the polarization fraction per radius. This suggests that the early-time dynamical response of the ring is characterized by coupled, oscillatory behavior of width and field orientation, whereas the polarization fraction reflects a delayed or modulated response, likely tied to changes in field ordering and geometry.

The integrated flux density $F_{\nu}$ of the entire simulation volume along a particular line of sight for a fixed source redshift $z$ is calculated numerically as the sum of the Stokes parameter $I$ of each grid cell multiplied by the solid angle subtended by the cell at the angular size distance at the given $z$.

For the fiducial Mach 4 shock case, the flux density evolution exhibits four distinct phases as shown in the last panel of Fig.~\ref{fig:radialplots}.
Initially, the flux density curve increases dramatically due to the compression by the shock-bubble interaction. After the shock passage, adiabatic expansion of the bubble reduces the electron density and magnetic field, causing $F_{\nu}$ to drop and remain roughly constant over time.

\subsubsection{Evolution of the Magnetic Field}
The total magnetic field energy, denoted as $B^2_{\text{strength}}$, is the sum of the magnetic energy density $u_{\rm B}$ of each grid cell multiplied by the volume $V$ of each grid cell, shown in the bottom panel of Fig.~\ref{fig:radialplots}. The figure shows a strong evolution of $B^2$. In the fiducial case, the magnetic energy shows a substantial initial decrease as the shock crosses the bubble. We interpret this as a result of numerical magnetic dissipation of the smallest scale modes during the shock compression, which compresses barely resolved changes in magnetic field orientation below the grid scale, as well as work done by the B-field to relax.

Numerical dissipation is an unavoidable consequence of non-spectral numerical simulations on Eulerian grids (needed to simulate compressible fluid mechanics.) We comment further on the role of numerical resolution on capturing the small evolution of the magnetic energy in \S\ref{subsec:differentresols}.

After about 10 shock crossing times, the magnetic energy reaches a minimum and then starts to increase again. We interpret the late time increase in $B^2$ as shear amplification in the differentially rotating vortex.

Importantly, this shear amplification process naturally explains the development of the radial magnetic field components observed in our synthetic polarization maps and statistics. As the vortex ring evolves, the differential rotation and shear flows systematically stretch and reorganize the magnetic field, preferentially amplifying field components aligned with the shear direction. This leads to the characteristic radial field patterns in the early and later stages of evolution and is reflected in both the polarization angle distributions and the depolarization effects we measure.

While the final field strength in our simulation is smaller than the initial field strength, likely due to numerical dissipation (see \S\ref{subsec:differentresols}), the dynamo amplification of the field during the development of the vortex ring is clearly visible in Fig.~\ref{fig:differentmachs}.
\begin{figure*}
    \centering
    \includegraphics[width=\linewidth]{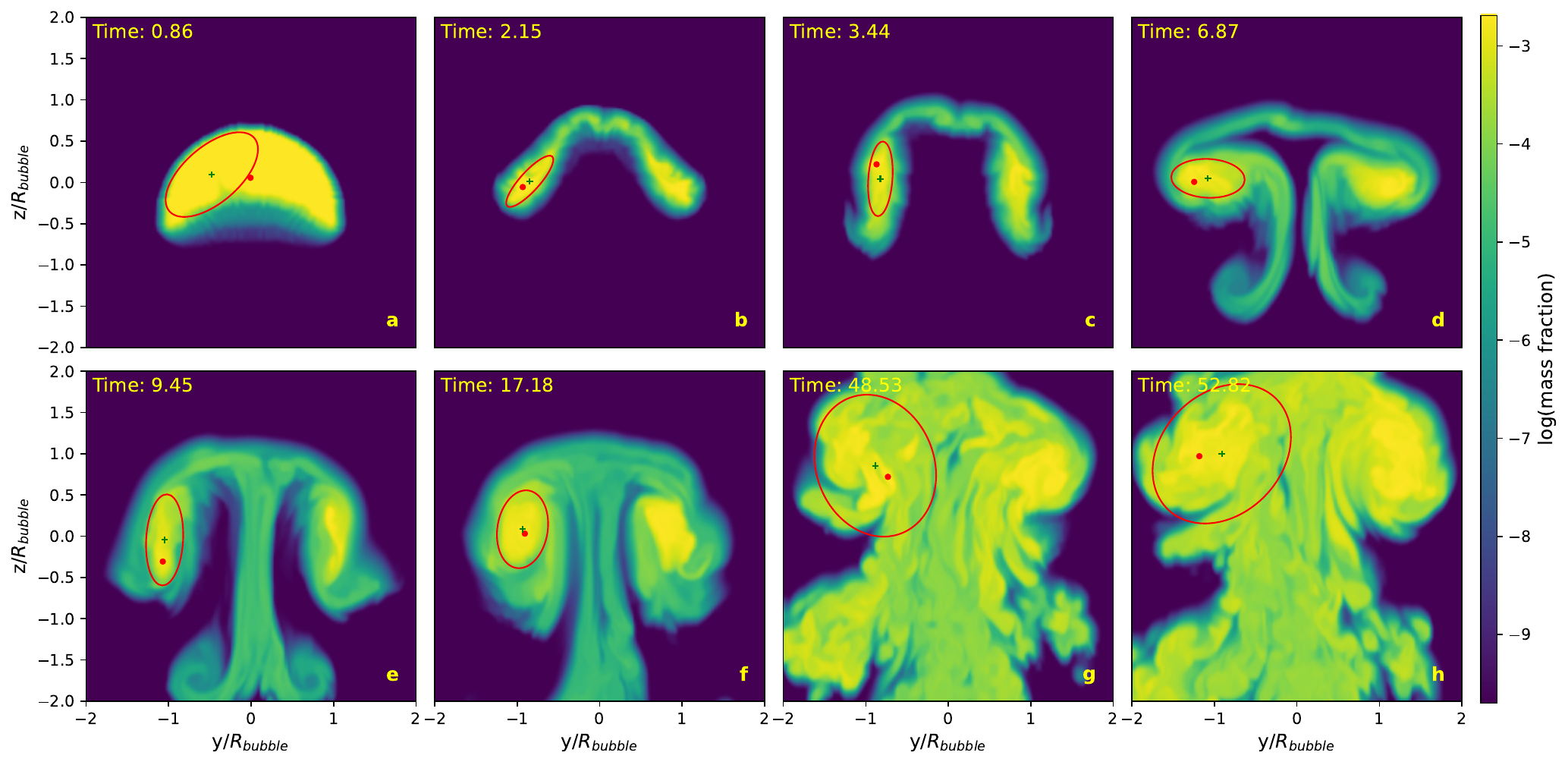}
    \caption{Slice plots in the x-direction (perpendicular to the shock normal and the line of symmetry of the vortex ring) of mass fraction of the non-thermal tracer fluid for the fiducial Mach 4 simulations, with best fit ellipses from the moment analysis of the images shown times $t=0.86, 2.15, 3.44, 6.87, 9.45, 17.18, 48.53, 52.82 t_{\rm cross}$, respectively.}
    \label{fig:Sliceplot}
\end{figure*}

\subsubsection{Transient Oscillatory Behavior}
In order to understand the initial strong oscillations in the observational quantities of the vortex, we analyze slice plots in a plane perpendicular to the image plane shown in Fig.~\ref{fig:Intensity}. Fig.~\ref{fig:Sliceplot} presents slice plots of the mass fraction of non-thermal gas within the bubble, focusing on the regions of interest around the left half of the vortex. The red point marks the location of the maximum mass fraction in the left half of the image. 

The first and second-order image moments are used to analyze the spatial distribution and dynamics of the non-thermal gas in these slices. First-order image moments are used to determine the centroid, or center of mass, of the non-thermal gas distribution. This is calculated by weighting each pixel's position by its fraction. The result is a green point representing the average location of the non-thermal gas in the left half of the image. 

Second-order image moments describe the extent and orientation of the gas distribution by calculating the semi-major and semi-minor axes of the ellipse best representing the data. 

These moments are used to fit an ellipse around the maximum mass fraction region (marked by the red point) in the left vortex. The turning of the ellipse around the vortex center shows the gas distribution revolving at the same frequency as the breathing motion observed in the intensity and polarization fraction.  That is, the vortex ring does not start out fully formed with a quasi-circular cross section. Instead, its cross section is elongated in a direction perpendicular to the ring itself. 

This deformation and revolution of fluids were first observed in \citep{haas_interaction_1987}. As the torus-like ring forms in the simulation, it exhibits a “breathing” motion: the projected vortex ring periodically contracts and expands in radius and thickness during the early evolution. A corresponding oscillation is observed in the ring's projected shape; as the ellipse rotates, the aspect ratio of the ring changes. We find that when the ellipse is more radially extended, corresponding to a smaller aspect ratio, the inferred magnetic field angle is larger, a result consistent with our analysis of the polarization morphology.

The transient nature indicates that vortex ring formation in shock-bubble interaction requires several turn-over times to set up a stable, converged vortex ring with quasi-circular cross section. It therefore carries important implications for the ability to constrain the age of a shock-driven vortex ring from the observational properties of ORCs which we will discuss further in \S\ref{sec:discussion}.

\subsection{Dependence on shock strength}\label{subsec:differentmachs}

\begin{figure}
\includegraphics[width=\linewidth]{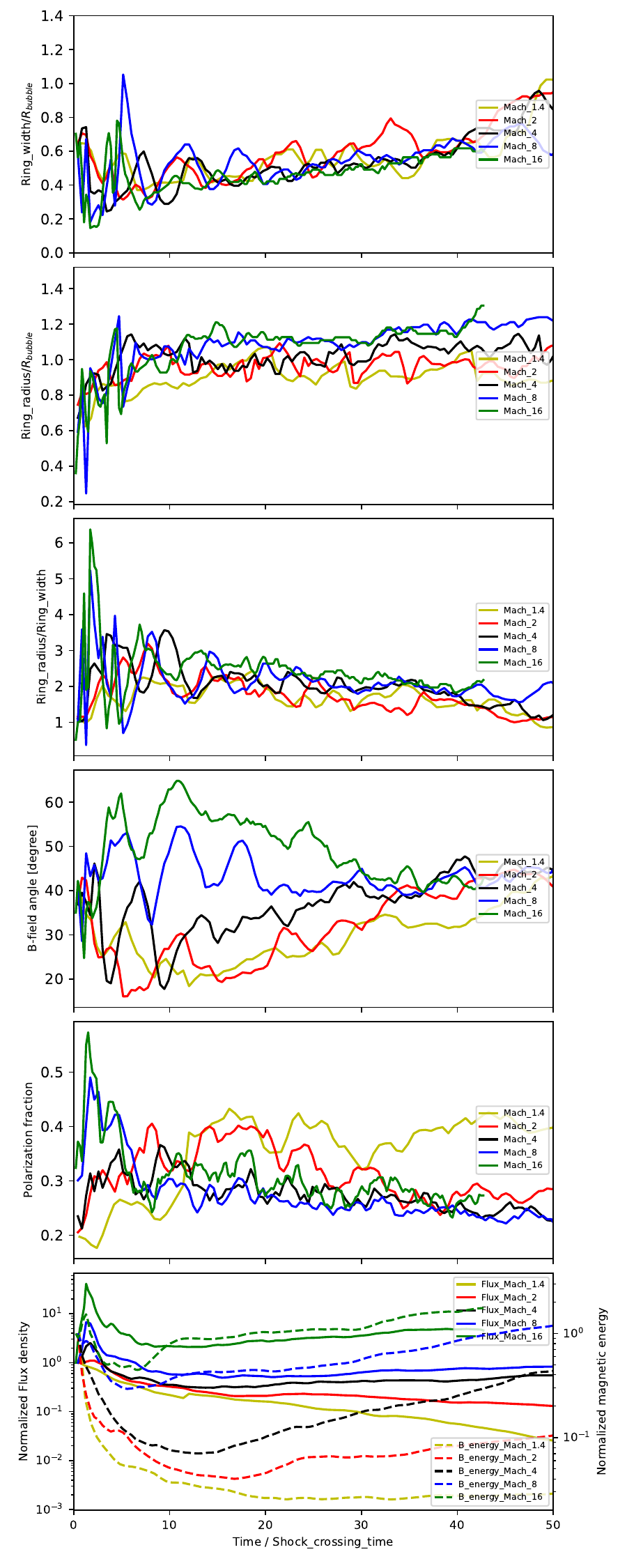}
\caption{Comparison of the same diagnostics shown in Fig.~\ref{fig:radialplots} for simulations with different Mach numbers of $M=1.44, 2, 4, 8, 16$ in yellow, red, black, blue, and green, respectively.
\label{fig:differentmachs}}
\end{figure}

To evaluate the dependence of our results on shock strength, we compare the five integrated observational properties of the bubble shown in Fig.~\ref{fig:radialplots} for simulations of shock-bubble interaction with varying Mach numbers otherwise identical to the fiducial Mach 4 case. The results are shown in Fig.~\ref{fig:differentmachs}.  

Once again, the first panel shows the width of the ring changing in time. The width consistently grows over time in every scenario, but the variability changes with Mach number. 

The evolution of the ring widths initially exhibits only moderate variations with Mach number. At the beginning, the oscillations for different Mach numbers peak at similar times, exhibiting similar phase and temporal patterns. The lower Mach numbers, however, eventually deviate from the higher Mach number cases: The Mach 1.4 and Mach 2 shocks generate wider, less well-defined rings at later times (we use the dotted lines to exclude the data where the ring cannot be identified), while the ring widths of the larger Mach number simulations remain relatively similar and show smaller overall amplitudes of fluctuation.  

The second panel displays the evolution of the ring's radius. For all Mach numbers, the simulations exhibit a progressive increase in radius over time, with the rate of growth being influenced by the corresponding Mach number such that lower Mach numbers result in a slower ring radius expansion and smaller ring radii.

The third panel presents the aspect ratio. Excluding the data where the ring can no longer be identified by eye in low Mach number cases (indicated by dotted lines), the raw (unsmoothed) aspect ratios of all our simulations are larger than the ratio reported in the literature\citep{norris_meerkat_2022}, i.e., the rings are narrower and more well-defined than the observed ORCs. However, this is largely a resolution effect, as we will discuss in \S\ref{sec:discussion}

The fourth panel once again shows the emission-weighted mean magnetic field angle measured relative to the tangent vector of the ring. In particular, we find that the Mach number has a strong effect on the observable polarization statistics during the formation and early evolution of the vortex, with higher Mach numbers showing a more radial magnetic orientation.  After strongly varying during the formation and initial evolution of the vortex ring, the mean angles converge to around 45 degrees at late times.

The first, third, and fourth panel show the same clear oscillations during the early period of vortex formation. The oscillations are correlated, and panels 3 and 4 are 180 degrees out of phase, i.e., in all cases, the field orientation shows the same anti-correlation to ring width as the Mach 4 case. 

The existence of phases with mostly radial magnetic fields in these vortex rings may provide a strong distinguishing observational diagnostic compared to other potential models of ORC formation.

The fifth panel presents the time evolution of the polarization fraction for shocks with different Mach numbers. At early times, all runs show strong correlation between the polarization fraction and their perspective B-field angle. In particular, the cases with higher polarization fractions coincide with moments when the B-field angle is smaller, i.e. more tangential. Conversely, when the B-field angle is larger, the polarization fraction drops. However, as the systems evolve, high Mach number shocks (Mach greater than 4), the polarization fraction steadily decays and converges toward a nearly constant value of ~25\%. Lower Mach number runs (Mach 1.4–2) maintain higher polarization fractions over longer timescales, reflecting weaker turbulence and are more diffusive in appearance.

The bottom panel shows the time evolution of the flux density (solid lines) and magnetic energy (dashed lines). The flux density exhibits a sharp initial rise due to the sudden compression of the gas and the magnetic field at the shock front, followed by a rapid decline, with the magnitude of the initial jump increasing with Mach number because stronger shocks generate higher post-shock compression. The magnetic energy shows only a modest early enhancement in the Mach 8 and Mach 16 cases, as the initial strong compression also amplifies the magnetic field component perpendicular to the shock. While in the lower Mach number cases, this compressive amplification is too weak, and magnetic energy initially decreases due to the volume expansion after the shock.
In all cases, the magnetic energy gradually rises again at later times. This late-time increase again reflects shear-driven magnetic amplification, which also contributes to the corresponding slow recovery in flux density.

\subsection{Dependence on Magnetic Topology}
\label{subsec:differentcuts}

To investigate the sensitivity of our results on the initial magnetic field configuration, 
we ran simulations with six different initial magnetic topologies by varying the power spectral parameters used to initialize the vector potential. 

\begin{figure}
\includegraphics[width=\linewidth]{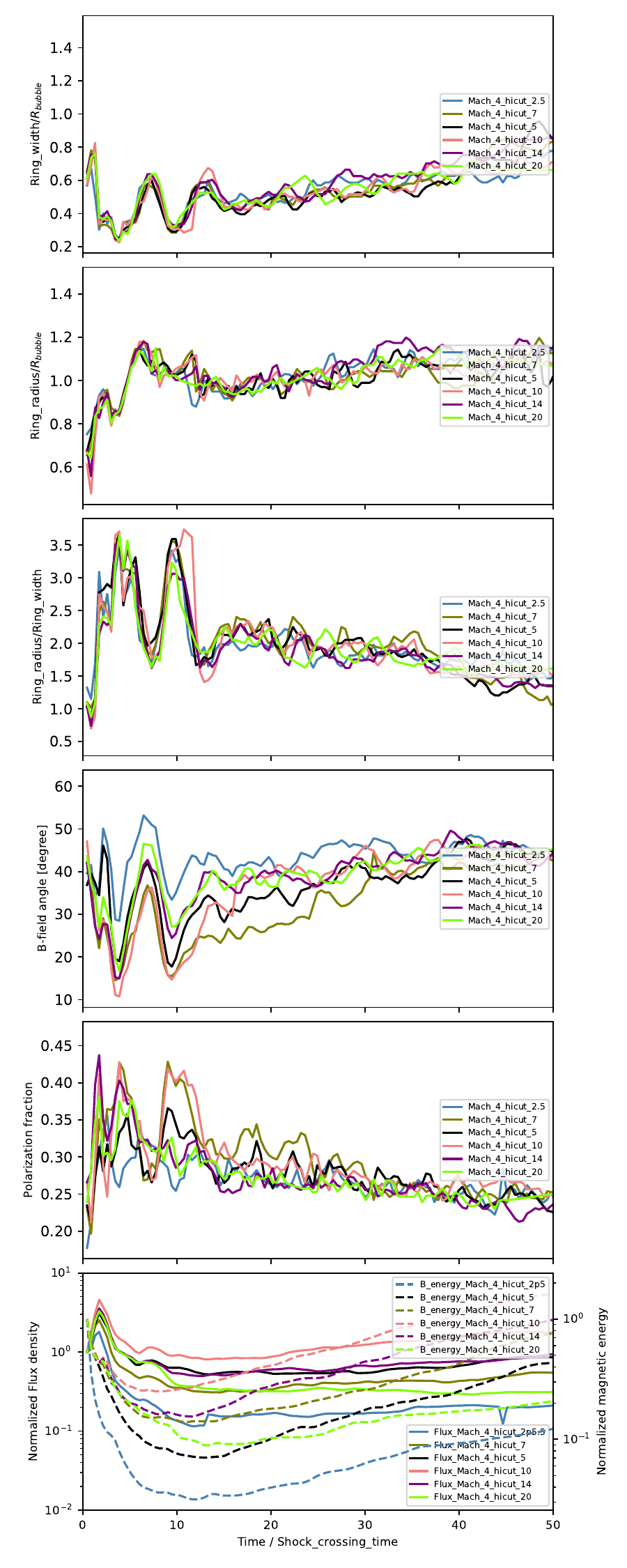}
\caption{Parameter comparison between different simulations with different magnetic topology and otherwise identical setup to the fiducial Mach 4 simulation. Shown are simulations with long-wavelengths cuts (characteristic coherence lengths) of 1/8, 1/4, 7/20, 1/2 (fiducial run), 7/10, and 1 bubble diameters and lower wavelength cutoffs (smallest scale of magnetic disorder) of half those values in blue, black, gray, red, purple, and green, respectively.}
\label{fig:differentcuts}
\end{figure}

Large cutoffs mean that the initial magnetic field coherence length is large, and the corresponding wave number $k$ is small. 
As shown in Figure \ref{fig:differentcuts}, the initial magnetic field configuration does not significantly affect the physical observables such as the ring width, ring radius, B-field angle, and polarization fraction. We thus interpret our findings relating to those observables as robust against magnetic topology.

However, magnetic topology does alter the magnetic field energy and flux density, shown in the bottom panel. Simulations with longer magnetic coherence length generate a larger flux density and magnetic field strength, possibly because the longer wavelengths involved are subject to less numerical dissipation.

However, when the magnetic coherence length is larger than the radius of the bubble, the initial magnetic field lines are mostly aligned inside the initial bubble, resulting in a decrease in the magnetic field strength and flux density. We must therefore be cautious regarding total flux density estimates against magnetic topology.  Our comparison with observational data in \S\ref{sec:discussion} is designed to be agnostic about the plasma beta and thus the absolute value of the magnetic energy.

In all cases, the simulations show late time shear amplification of the magnetic field in the differentially rotating vortex.

We verified that variations in the random seeds used to initialize the magnetic field are found not to significantly impact the physical properties of the simulated rings, provided the same power spectral parameters are otherwise maintained.

\subsection{Resolution Study}
\label{subsec:differentresols}

All simulations discussed so far have identical resolution as the fiducial Mach 4 simulation, i.e., a resolution of 128 cells across the bubble diameter. 

\begin{figure}
\includegraphics[width=\linewidth]{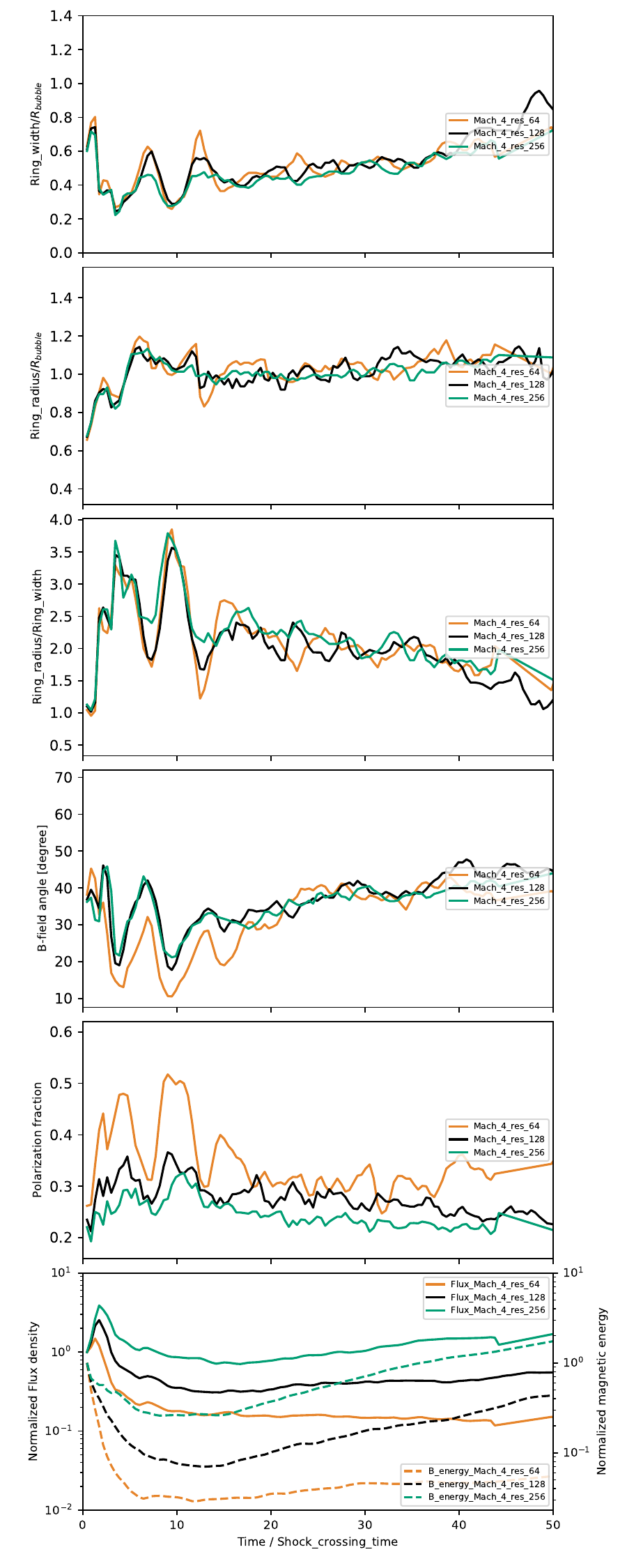}
\caption{Parameter comparison between different resolutions, showing the same diagnostics as a function of time as Fig.~\ref{fig:radialplots} for simulations with resolutions of 64 cells, 128 cells, and 256 cells across the bubble in orange, black, and green, respectively
\label{fig:differentres}}
\end{figure}

In order to evaluate which of our diagnostic quantities and conclusions are robust against changes in resolution, or have converged at the nominal resolution of 128 cells, we ran two additional simulations of our fiducial Mach 4 case with resolutions 64 and 256, respectively. The results of the same diagnostics as in Fig.~\ref{fig:radialplots} are shown in Fig.~\ref{fig:differentres}.

We find that changes in the simulation resolution do not significantly affect the ring's width and radius, and we can consider these quantities as robust against resolution changes. 

At early stages, the magnetic field angle in the lowest resolution case is smaller than the higher resolution cases but asymptotically approaches the same 45 degree angle at late times. We consider the angle diagnostic to be converged at our nominal resolution of 128 cells across a bubble diameter.

The polarization fraction decreases with increasing resolution. This is not unexpected, as higher numerical resolution will lead to increasing in-beam depolarization as smaller field reversals can be captured in the line-of-sight integrals. We therefore consider the predicted polarization fractions of runs conducted at our fiducial resolution as upper limits, however, we expect net polarization to remain well above detection thresholds as resolution is increased.

Additionally, we find a strong dependence of magnetic field strength and flux density on resolution. The field strength increases by almost an order of magnitude as we increase resolution from 64 to 256 cells across the bubble, while the flux density roughly doubles with each doubling of the resolution.  This confirms our interpretation that magnetic field evolution and flux estimates suffer from numerical dissipation effects. 

Higher-resolution simulations can resolve smaller-scale modes, effectively capturing shorter wavelengths, and are less prone to numerical reconnection. Conversely, lower-resolution simulations tend to have higher numerical dissipation, which can artificially smooth out magnetic field variations and thus increase numerical magnetic reconnection rates.  This smoothing can create a more uniform magnetic field, leading to a higher polarization fraction, as can be seen in panel five.

We thus conclude that neither flux density nor total intensity (both observables) are converged at our nominal resolution, which is entirely due to the plasma beta not being converged. Similarly, magnetic energy (not observable) is also not converged in runs at our fiducial resolution. However, because the dependence of flux density on the magnetic field strength (and thus plasma beta) is well understood, our scale free approach to calculating synthetic radio maps allows us to leave the total field strength a free parameter in model fitting in \S\ref{sec:discussion}, circumventing the lack of convergence of these particular parameters when making comparisons with observations..

We find that the {\em relative} increase in magnetic energy at late times is consistent between our fiducial resolution of 128 cells and the high resolution simulation, indicating that the shear amplification of the field is well captured by simulations at the nominal resolution of 128.

\section{Discussion}\label{sec:discussion}

\subsection{Comparison With Observations and Implications}
Table~\ref{tab:ORCs} summarizes the observational data for the sample of known ORCs. For the initial discussion of our results, we limit our comparison with these data to our Mach 2 and Mach 4 simulations, as they provide the most relevant physical and morphological analogues. Cases with Mach $>$ 4 are excluded because such strong shocks are not expected to form during large-scale cosmic structure formation and are exceedingly rare in large scale astrophysical contexts. Furthermore, the Mach 1.4 simulation is excluded because we cannot extract a reasonable peak or FWHM from its intensity profile, indicating a ring structure that is too diffuse to match the well-defined morphology of observed ORCs.

For direct comparison with the observational data, we convolved images with a Gaussian kernel of the telescope's beam width (FWHM) for the ASKAP image of ORC5 (13 arcsec for an ORC radius of 35 arcsec, or 0.37 beams per radius). The resulting rings show similar radius but larger ring width, decreasing the aspect ratio as shown in Fig.~\ref{fig:ratiomatch}.

\begin{deluxetable*}{c|c|c|c|c|c|c|c|c}
\tablecaption{Observed ORCs and properties assuming the redshifts of the central galaxies. Shown are the literature identification number ordered by date of discovery (left column), the associated object name (second column), the telescope used for discovery (third column), the image resolution (fourth column), the redshift of the central galaxy (fifth column), the ring radius (sixth column), the integrated flux density (seventh column), the observing frequency (eighth column), and the spectral index (last column). Superscript symbols indicate references for the listed properties: 
($^{\dagger}$) \citet{norris_unexpected_2021}, 
($^{\star}$) \citet{koribalski_discovery_2021}, 
($^{\ddagger}$) \citet{norris_meerkat_2022}, 
($^{\mathsection}$) \citet{lochner_unique_2023}, 
($^{\|}$) \citet{koribalski_meerkat_2024},
($^{\#}$) \citet{bulbul_galaxy_2024}.
\label{tab:ORCs}}
\tablewidth{0pt} 
\tablehead{
\colhead{ORC} & \colhead{Name} & \colhead{Telescope} & \colhead{Resolution} & \colhead{Galaxy redshift} & \colhead{Radius} & \colhead{Integrated flux} & \colhead{Frequency} & \colhead{Spectral Index} \\
\colhead{Number} & \colhead{} & \colhead{} & \colhead{(arcsec)} & \colhead{ (z)} & \colhead{(kpc)} & \colhead{(mJy)} & \colhead{ (MHz)} & \colhead{$\alpha$} }
\startdata
ORC1$^{\dagger}$   & J210357 & ASKAP   & 11          & 0.55  & 255   & 6.26$\pm$ 1.25 & 944   & -1.17$\pm$ 0.04 \\
ORC2$^{\dagger}$   & J205842 & ASKAP   & 13          & 0.311 & 260   & 6.97$\pm$1.39 & 944   & -0.80$\pm$ 0.08 \\
ORC3$^{\dagger}$   & J205856 & ASKAP   & 13          &  --   & --    & 1.86$\pm$ 0.37 & 944   & -0.50$\pm$ 0.2 \\
ORC4$^{\ddagger}$  & J1555   & GMRT    & 9.4$\times$7.9 & 0.39  & 185   & 28$\pm$ 2.8   & 325   & -0.92$\pm$ 0.18 \\
ORC5$^{\star}$     & J0102   & ASKAP   & 13          & 0.27  & 150    & 3.9  & 944   & -0.80$\pm$ 0.2 \\
ORC6$^{\|}$        & J1027   & MeerKAT & 7.7         & 0.3   & 200   & 1.2$\pm$ 0.1 (0.35) & 888(1600) &  -2.10$\pm$ 0.5 \\
Possible ORC$^{\mathsection}$ & SAURON & MeerKAT &--  & 0.55 & 185 &   3.9 &1400    &  -1.4 \\
\textcolor{gray}{ORC7$^{\#}$} & \textcolor{gray}{Cloverleaf} & \textcolor{gray}{ASKAP} & \textcolor{gray}{20} &  \textcolor{gray}{0.05}  &  \textcolor{gray}{227$\times$137} & -- & \textcolor{gray}{856} & \textcolor{gray}{-1}\\
\textcolor{gray}{X-ray$^{\#}$} & \textcolor{gray}{Cloverleaf} & \textcolor{gray}{XMM-Newton} & -- & -- & \textcolor{gray}{100$\times$180} & -- &  -- &  --\\
\enddata
\end{deluxetable*}

\subsubsection{Aspect Ratio (Radius over Width) and size}
\begin{figure*}
    \centering
    \includegraphics[width=\linewidth]{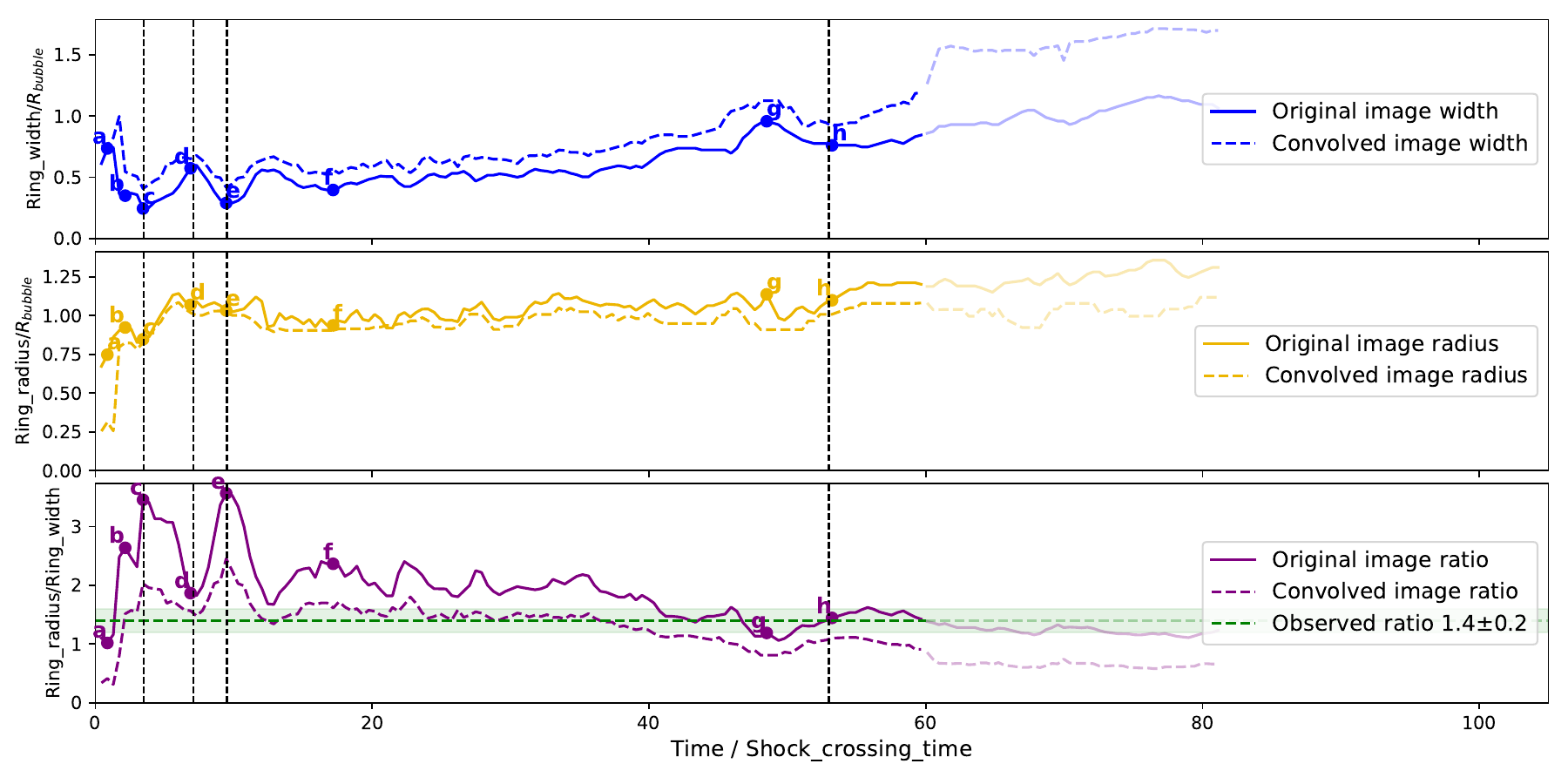}
\caption{Observational diagnostics compared to ORC5, following the top three panels of Fig.~\ref{fig:radialplots} for raw (solid lines) and smoothed data (dashed lines). There's no well-defined rings/circles after t/t$_\text{cs}$ = 60 so we do not consider the data after that. This provides a loose constraint on the age of the circles.}
    \label{fig:ratiomatch}
\end{figure*}

The aspect ratio of a synthetic radio image of the vortex ring, defined as the ratio of the ring radius to its width, is a robust and easily measured diagnostic. We will base our analysis in this section on comparisons with observations of ORCs. Due to the lack of published quantitative measurements of aspect ratios for other ORC sources, we have used the aspect ratio value from ORC5 J0102 with a published ratio of 1.4, as our reference point throughout this analysis (with individual ORC radii summarized in \S\ref{tab:ORCs}). Figure \ref{fig:ratiomatch} shows the ring width, ring radius and the aspect ratio (the third panel) from the convolved images of our Mach 4 case. We find that the simulated rings have larger aspect ratios than the observed ORC5. However, the low resolution of the radio images of ORCs in general suggests that we must compare quantities after smoothing. We can see from Fig.~\ref{fig:ratiomatch} that the aspect ratio of simulated rings of our fiducial case generally agrees well with the observation once resolution effects are taken into account at early times, with simulated aspect ratios only marginally larger than the observed value.

For any given simulation time, we can then find the matching set of length scale parameters by matching the observed ring radius to the simulated (smoothed) one: 
\begin{equation}
    R^{\text{obs}}_{bubble}= \frac{R_\text{ring}^{\text{obs}}}{R_\text{ring}^{\text{sim}}}R^{\text{sim}}_{bubble}
\end{equation}

In order to reproduce the size of the observed ring from the shock-bubble interaction, we infer an initial bubble radius of order 150 kpc for ORC5 modeled by our Mach 4 simulation. And ORC1 and ORC2 would require even larger initial bubbles. The required initial bubble radius results for all ORCs with the Mach 2 and 4 cases are summarized in Table 3, assuming the published redshifts of the center-most galaxy. This initial bubble size scale is consistent with the radii of active or fossil radio lobes of powerful radio galaxies, which we suggest are the parent population of ORCs, consistent with \citep{shabala_are_2024}.

To put this into physical context, we note that jet-inflated bubbles in galaxy clusters have sizes ranging from tens to hundreds of kiloparsecs within the intracluster medium (ICM). For example, the Perseus Cluster contains cavities about 20-30 kpc in size \citep{Fabian_2006}. But extremely energetic outbursts of cluster radio galaxies, such as in MS 0735.6+7421 \citep{vantyghem_cycling_2014}, can generate cavities of about 200 kpc. The Cygnus A lobes span about of 140kpc \citep{emonts_large-scale_2007} and the Centaurus A lobes has approximately 650kpc in diameter \citep{clarke_vla_1992}. Similarly, radio lobes of radio loud AGN in field galaxies can extend over hundreds of kiloparsecs, and buoyant expansion can further increase their effective cross section. Our model thus aligns well with the properties of powerful AGN. We note such large fossil bubbles imply an origin in extreme radio-galaxy activity or group-scale outflows and are therefore expected to be rare. Smaller fossil bubbles are expected to produce radio rings that are both fainter and shorter-lived. Their thinner projected width, faster radiative cooling, and reduced surface brightness would make them difficult to identify as ORCs in current surveys. As a result, observational selection effects strongly bias ORC detections toward the largest and most energetic fossil structures, providing a natural explanation for both the large inferred bubble sizes and the apparent rarity of ORCs.

\subsubsection{Constraints from Model Grids with Unconstrained Redshifts}
\label{sec:discussion_modelfitting}
In addition to the ring radius, which we can use to determine one of the simulation scale parameters (the initial bubble radius) we can also use the flux density and spectral slope to constrain the remaining parameters.

A key distinguishing feature of our model, relative to most other prospective mechanisms of ORC formation, is that it is not inherently tied to the redshift of any central galaxy; the shock-bubble interaction mechanism itself is redshift-agnostic. This is because the host galaxy responsible for inflating the bubble does not have to be aligned with the center of the bubble, depending on the orientation of the jet axis at the time of inflation. To fully leverage this and explore the complete parameter space, we initially treat the redshift as a free parameter. This allows us to determine the range of physical conditions under which an ORC-like object could be produced, independent of a specific host identification.

Given the observing frequencies and the likely ages of the synchrotron plasma in ORCs, it is critical to properly account for radiative losses in the calculation of flux densities for comparison with the observational values.

We thus follow the calculation laid out in \S\ref{sec:synchrotron} and compare our values to the comoving physical parameters (i.e., adjusted for cosmological dimming and red-shift.)

We built a 6-D grid of model predictions for each ORC over $(M, f_{\text{n}}, z, \beta P, T)$, where $M$ is the shock Mach number, $f_n$ indexes simulation snapshot file, $z$ is redshift, $\beta$ is the plasma beta, $(P,T)$ span a pressure-temperature range. For each grid point we record the predicted integrated flux at the observed frequency, the local spectral index $\alpha$, the thermal and magnetic energies,the source age, a chi-square statistic $\chi^2$, and a likelihood $L=\exp(-\chi^2/2)$. The grid comprises

\begin{itemize}
\item{Shock strengths with Mach numbers 2, 2.83, and 4}
\item{Redshifts spanning the range $z \in[0.1,1.0]$, sampled logarithmically with $n_z=20$ grid points}
\item{Plasma betas for in the range $\beta \in\left[0.1, 10^3\right]$ sampled logarithmically with $n_\beta=20$ grid points}
\item{IGM pressures spanning $P \in\left[10^{-16}, 10^{-11}\right] \mathrm{dyn} \mathrm{ cm}^{-2}$, logarithmically sampled on a $200$ points}
\item{IGM temperatures in the range $T \in\left[10^6, 10^8\right] \mathrm{K}$, logarithmically sampled on a $200$ points}
\item{21 snapshots in time ($f_{\text {n }}$) from each simulation series from 2.15 t$_\text{cross}$ to 25.78 t$_\text{cross}$.}
\end{itemize}
spanning frequency  from 0.3 MHz to 800000.0 MHz. 
We impose a minimal ICM density prior of $n > 5 \times 10^{-7}$, corresponding to the cosmic mean, on this grid. The pressure range is chosen arbitrarily since no rich clusters have been found in or near the observed ORCs\citep{norris_unexpected_2021}.

Once again, we follow the calculation laid out in \S\ref{sec:synchrotron}, computing the synthetic synchrotron spectra for each of the two loss mechanisms (synchrotron and inverse Compton) separately. We then combine them with a "parallel sum" to calculate the actual spectrum, which is approaches the smaller of the two when one channel dominates:
\begin{equation}
    F_{\nu, \mathrm{comb}}=\left(\frac{1}{F_{\nu, \mathrm{IC}}}+\frac{1}{F_{\nu, \mathrm{syn}}}\right)^{-1}
\end{equation}
The effective spectral index is the flux-weighted $\alpha_{\mathrm{comb}}$
\begin{equation}
    \alpha_{\mathrm{comb}}=\frac{F_{\nu, \mathrm{IC}} \alpha_{\mathrm{IC}}+F_{\nu, \mathrm{syn}} \alpha_{\mathrm{syn}}}{F_{\nu, \mathrm{IC}}+F_{\nu, \mathrm{syn}}}
\end{equation}
At each grid point we compare to the observed flux and spectral index. Our goodness of fit parameter is
\begin{align*}
    \chi^2=&\left(\frac{F_{\nu, \mathrm{obs}}-F_{\nu, \mathrm{comb}}}{\sigma_F}\right)^2+\left(\frac{\alpha_{\mathrm{obs}}-\alpha_{\mathrm{comb}}}{\sigma_\alpha}\right)^2\\
    & +\left[\frac{\max \left(0, \rho_{\min }-\rho_{\mathrm{bubble}}\right)}{\sigma_\rho}\right]^2
\end{align*}
and the likelihood function is
\begin{equation}
    L = \exp (-\chi^2/2)
\end{equation}
We calculate the 3D likelihood map for each simulation output file to generate likelihood contours as functions of time. To generate confidence intervals for individual parameters or parameter pairs, we can marginalize over the remaining parameters.
%

\begin{figure*}[ht]
\centering
\hspace*{-1.7cm}
\includegraphics[width=\textwidth]{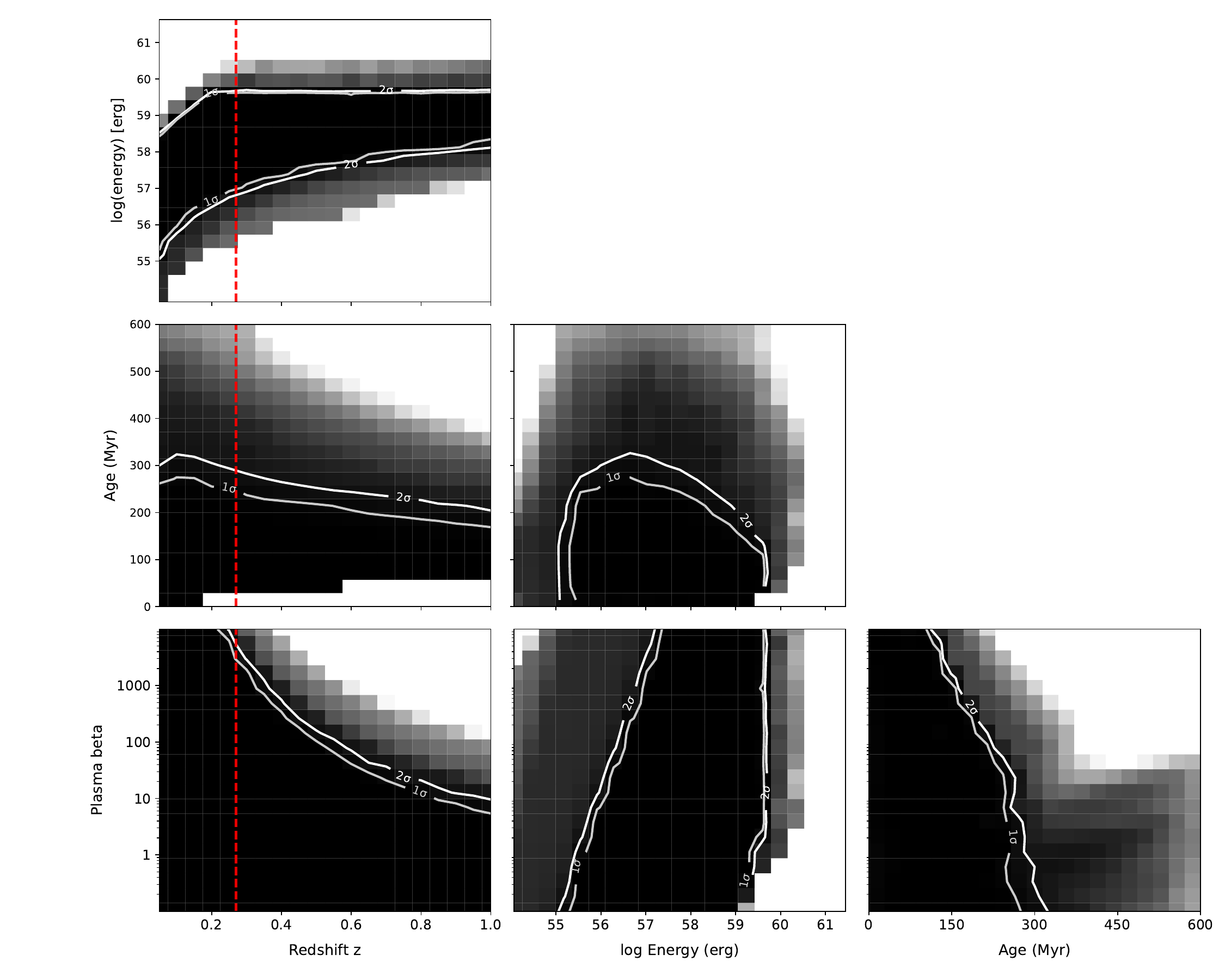} 
\caption{Corner plots for two-dimensional likelihoods functions for a wide range of model parameters for Mach4–ORC5. 
Each panel shows the likelihood map and contours (maximum over the remaining parameters) for the indicated pair, with grayscale shading (black = highest likelihood) and white contours at $\Delta\log L$ = -1.15, -3.09 ($1\sigma, 2\sigma$). 
Panels, read left-to-right, top-to-bottom: 
(0,0) redshift z vs. log (energy/erg); 
(1,0) redshift z vs. age (Myr); 
(1,1) log (energy/erg) vs. age (Myr); 
(2,0) redshift z vs. plasma $\beta$ (y-axis logarithmic); 
(2,1) plasma $\beta$ (log) vs. log (energy/erg); 
(2,2) plasma $\beta$ (log) vs. age (Myr). 
Energy is binned uniformly in log10, age linearly in [0,100] Mry. Diagonal panels are omitted, and axes are labeled with the corresponding parameters: redshift $z$, log(energy) in erg, age in Myr, and plasma $\beta$. The red dashed  vertical line indicates the redshift of the central galaxy associated with ORC5, showing a wide range of possible energy and plasma beta values.}
\label{fig:merged_plot}
\end{figure*}

Figure~\ref{fig:merged_plot} shows corner plots for the likelihood function of different parameter combinations. To generate them, we marginalized over all three Mach numbers Mach 2, the intermediate case (Mach 2.83), and Mach 4, and all other remaining parameters. As established previously.
 We then show the max likelihood values as a heatmap using a logarithmic color normalization 
so that a wide dynamic range can be displayed; darker tones (with the gray map) indicate higher likelihood. The likelihood shown here is computed using only those models that are consistent with the observed ORCs in both morphology and radio spectrum, since our goal is to identify physically plausible solutions that simultaneously reproduce these key observational constraints. Light regions indicate the range of parameters of our SBI models that ar incompatible with the observations because the predicted spectra are inconsistent with the measured flux densities.  We overlay $\Delta\log $-likelihood contours to indicate approximate $1\sigma$ and $2\sigma$ confidence regions.
\begin{itemize}
\item{Plot (0,0): $(z-\log E)$ shows the highest likelihood follows an increasing ridge of energy with redshift, i.e. larger z requires larger total energy. This is expected due to the luminosity-distance scaling.}
\item{Plot (1,0): $(z-t)$ shows that the likelihood is concentrated at younger than 200 Myr across the redshift range.}
\item{Plot (1,1): $(\log E -t)$ shows a modest degeneracy of energy and age, with a range of energies and times of $E \approx 10^{57}-{\rm few}\times 10^{59} $ erg and $t\approx 100-300$ Myr, respectively.}
\item{Plot (2,0): $(z-\beta)$ shows higher z values prefer lower $\beta$ (stronger $B$), while lower z allows a broad range of $\beta$ values.}
\item{Plot (2,1): $(\beta - \log E)$ shows that the total energy is largely independent of $\beta$, as it is set by the external pressure and bubble radius.}
\item{Plot (2,2): $(\beta-t)$ shows age is largely independent of $\beta$.}
\end{itemize}

\subsubsection{Model Constraints with Redshift Prior Fixed to the Central Source}

\begin{figure}
    \centering
    \includegraphics[width=1.25\columnwidth]{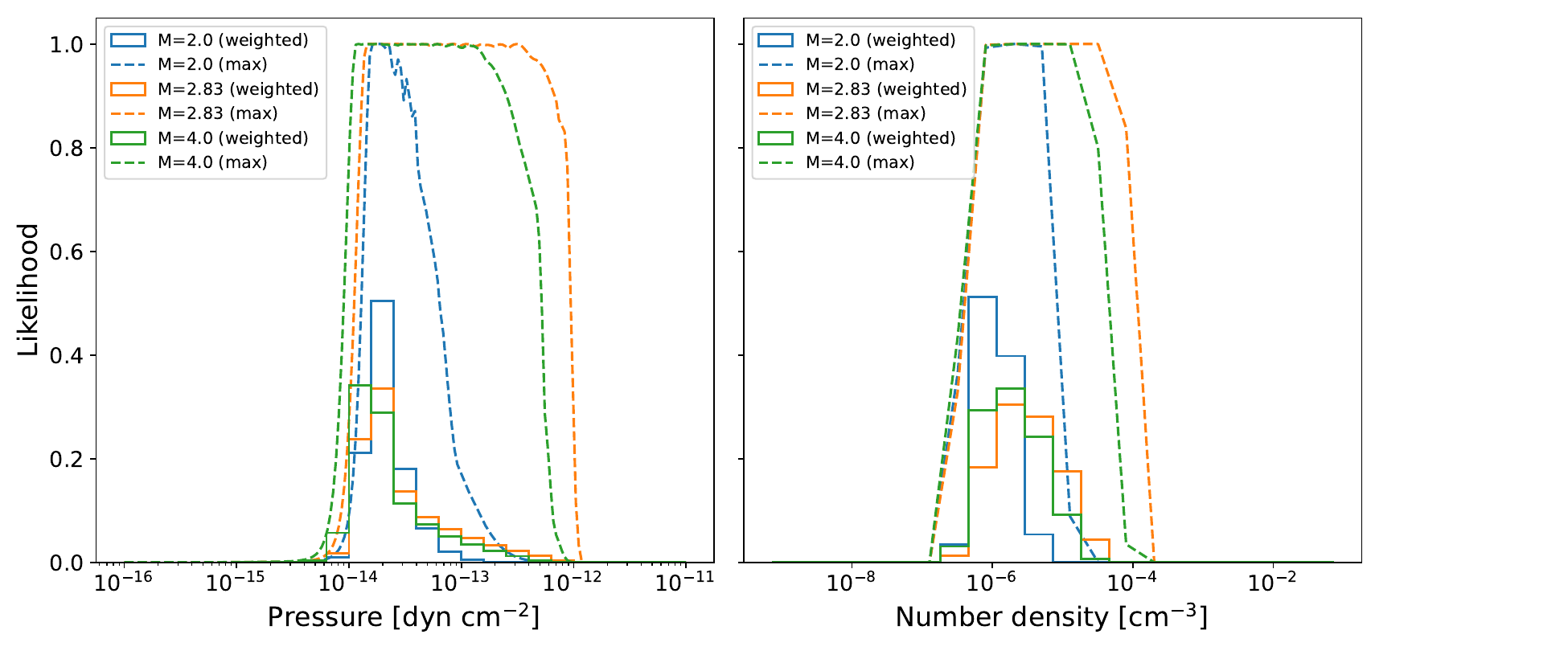}
    \caption{Likelihood distributions of \textbf{pressure} and \textbf{number density} for the ORC5 flux-matched models at 944~MHz, assuming Mach 2.0, 2.83, 4.0 shocks, and fixed z= 0.27. 
    Solid curves show the marginalized likelihood histograms (20~bins), and dashed curves indicate the one-dimensional profile-likelihood maxima normalized to unity. 
    The derived $1\sigma$ ranges from the profile-
likelihood maxima are: for Mach~2.0, $P = 1.4\times10^{-14}$--$5.5\times10^{-14}\,\mathrm{dyn\,cm^{-2}}$ and $n = 8.0\times10^{-7}$--$5.0\times10^{-6}\,\mathrm{cm^{-3}}$; 
    for Mach~2.83, $P = 1.2\times10^{-14}$--$8.8\times10^{-13}\,\mathrm{dyn\,cm^{-2}}$ and $n = 8.0\times10^{-7}$--$8.0\times10^{-5}\,\mathrm{cm^{-3}}$; 
    and for Mach~4.0, $P = 9.7\times10^{-15}$--$4.9\times10^{-13}\,\mathrm{dyn\,cm^{-2}}$ and $n = 8.0\times10^{-7}$--$3.2\times10^{-5}\,\mathrm{cm^{-3}}$. 
}
    \label{fig:pressuredensitydistribution}
\end{figure}

\begin{figure}
    \centering
    \includegraphics[width=1.25\columnwidth]{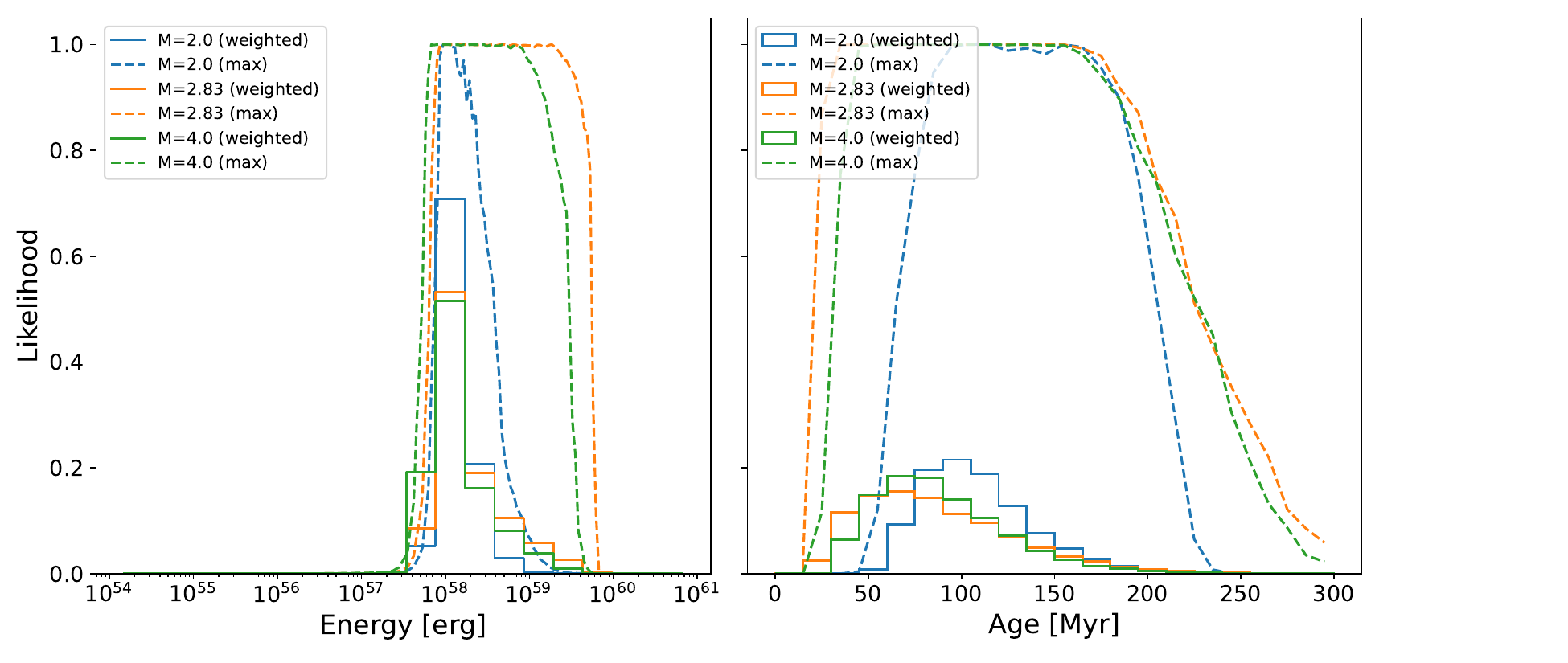}
    \caption{Likelihood distributions of \textbf{initial bubble energy} and \textbf{source age} for the ORC5 flux-matched models at 944~MHz, assuming Mach 2.0, 2.83, 4.0 shocks, and fixed z= 0.27. 
    Solid curves show the marginalized likelihood histograms (20~bins), and dashed curves indicate the one-dimensional profile-likelihood maxima normalized to unity. 
    The derived $1\sigma$ ranges from the profile-likelihood maxima are:
    for Mach~2.0, $E = 8.0\times10^{57}$--$3.1\times10^{58}\,\mathrm{erg}$ and $t = 75$--$195\,\mathrm{Myr}$;
    for Mach~2.83, $E = 6.8\times10^{57}$--$5.3\times10^{59}\,\mathrm{erg}$ and $t = 25$--$215\,\mathrm{Myr}$;
    and for Mach~4.0, $E = 5.8\times10^{57}$--$2.8\times10^{59}\,\mathrm{erg}$ and $t = 35$--$205\,\mathrm{Myr}$.
}
    \label{fig:ageenergydistribution}
\end{figure}

We now specialize to the case most readily comparable with current observations: the assumption that the central galaxy closest to the ring center is indeed the host. This imposes the galaxy's observed redshift as a prior, fixing the luminosity distance and allowing us to derive more precise estimates for the physical parameters of each source. Under this assumption, we obtain likelihood-weighted distributions for the bubble energy, ring age, and ambient conditions. 
Fig.~\ref{fig:pressuredensitydistribution} shows the normalized posterior distributions of pressure and number density derived from the likelihood analysis for ORC5. The dashed curves illustrate the full range of values that are consistent with the model once all other parameters are marginalized (except the fixed redshift), whereas the histograms represent the likelihood-weighted distributions, highlighting which parameter ranges are statistically more favored when accounting for the likelihood contributions of all other parameters. The 1-$\sigma$ range of the dashed curves is given as the interval where the likelihood remains within one standard deviation of the maximum value, indicating the probable parameter region. Fig.~\ref{fig:ageenergydistribution} similarly presents the normalized posterior distributions of initial total energy and source age.

The best-fit values of the initial bubble radius, ambient pressure, density, ring age, and total energy for all published ORCs are summarized in Tab.~\ref{tab:example}, using the results from Mach 2 and Mach 4 shock simulations as representative lower and upper limits, respectively.

\begin{deluxetable*}{cccccc}
    \tablecaption{Estimated initial bubble radius, ambient pressure, number density, ring age, and total injected energy for different Odd Radio Circles (ORCs) derived from weighted likelihood histograms, assuming the redshifts of central galaxies.
        Each ORC was analyzed using the Mach 2(M2) and Mach 4(M4) shock–bubble interaction cases.
        The pressure and density values show in this table represent the most likely ranges of the surrounding intergalactic medium, while a broader range of physically allowed parameters is indicated by the $1\sigma$ intervals in Fig.~\ref{fig:pressuredensitydistribution} and Fig.~\ref{fig:ageenergydistribution}. The ring ages correspond to the post-shock expansion time. The total energy corresponds to the initial energy of the bubble prior to shock compression. The intermediate Mach 2.83 case lies between the two limits and is omitted for clarity. The uncertainties represent half the range between the most likely lower and upper bins of each parameter and are therefore subject to the adopted binning resolution.
 \label{tab:example}}
    \tablehead{
        \colhead{ORC} & 
        \colhead{Initial bubble radius (kpc)} & 
        \colhead{Pressure (dyn cm$^{-2}$)} &
        \colhead{Density (cm$^{-3}$)} &
        \colhead{Ring age (Myr)} & 
        \colhead{Total initial energy (erg)}
    }
    \startdata
        ORC1 & 
        \begin{tabular}[c]{@{}l@{}}244.8 $\pm$ 32.3 (M2)\\ 239.1 $\pm$ 12.6 (M4)\end{tabular} &
        \begin{tabular}[c]{@{}l@{}}(3.25 $\pm$ 0.73) $\times 10^{-14}$\\ (3.25 $\pm$ 0.73) $\times 10^{-14}$\end{tabular} &
        \begin{tabular}[c]{@{}l@{}}(8.02 $\pm$ 3.45) $\times 10^{-7}$\\ (2.02 $\pm$ 0.87) $\times 10^{-6}$\end{tabular} &
        \begin{tabular}[c]{@{}l@{}}142.5 $\pm$ 7.5\\ 142.5 $\pm$ 7.5\end{tabular} &
        \begin{tabular}[c]{@{}l@{}}(1.29 $\pm$ 0.42) $\times 10^{59}$\\ (1.29 $\pm$ 0.42) $\times 10^{59}$\end{tabular} \\
        ORC2 & 
        \begin{tabular}[c]{@{}l@{}}252.8 $\pm$ 36.1 (M2)\\ 245.9 $\pm$ 10.7 (M4)\end{tabular} &
        \begin{tabular}[c]{@{}l@{}}(1.29 $\pm$ 0.29) $\times 10^{-14}$\\ (8.16 $\pm$ 1.85) $\times 10^{-15}$\end{tabular} &
        \begin{tabular}[c]{@{}l@{}}(3.20 $\pm$ 1.38) $\times 10^{-7}$\\ (8.02 $\pm$ 3.45) $\times 10^{-7}$\end{tabular} &
        \begin{tabular}[c]{@{}l@{}}142.5 $\pm$ 7.5\\ 97.5 $\pm$ 7.5\end{tabular} &
        \begin{tabular}[c]{@{}l@{}}(5.75 $\pm$ 2.20) $\times 10^{58}$\\ (2.57 $\pm$ 0.98) $\times 10^{58}$\end{tabular} \\
        ORC4 & 
        \begin{tabular}[c]{@{}l@{}}144.9 $\pm$ 19.9 (M2)\\ 141.9 $\pm$ 6.2 (M4)\end{tabular} &
        \begin{tabular}[c]{@{}l@{}}(5.15 $\pm$ 1.17) $\times 10^{-14}$\\ (5.15 $\pm$ 1.17) $\times 10^{-14}$\end{tabular} &
        \begin{tabular}[c]{@{}l@{}}(5.06 $\pm$ 2.18) $\times 10^{-6}$\\ (5.06 $\pm$ 2.18) $\times 10^{-6}$\end{tabular} &
        \begin{tabular}[c]{@{}l@{}}142.5 $\pm$ 7.5\\ 127.5 $\pm$ 7.5\end{tabular} &
        \begin{tabular}[c]{@{}l@{}}(1.29 $\pm$ 0.42) $\times 10^{59}$\\ (5.75 $\pm$ 2.20) $\times 10^{58}$\end{tabular} \\
        ORC5 & 
        \begin{tabular}[c]{@{}l@{}}166.7 $\pm$ 3.7 (M2)\\ 144.4 $\pm$ 5.8 (M4)\end{tabular} &
        \begin{tabular}[c]{@{}l@{}}(2.05 $\pm$ 0.46) $\times 10^{-14}$\\ (1.29 $\pm$ 0.29) $\times 10^{-14}$\end{tabular} &
        \begin{tabular}[c]{@{}l@{}}(8.02 $\pm$ 3.45) $\times 10^{-7}$\\ (2.02 $\pm$ 0.87) $\times 10^{-6}$\end{tabular} &
        \begin{tabular}[c]{@{}l@{}}97.5 $\pm$ 7.5\\ 67.5 $\pm$ 7.5\end{tabular} &
        \begin{tabular}[c]{@{}l@{}}(1.15 $\pm$ 0.43) $\times 10^{58}$\\ (9.00 $\pm$ 3.00) $\times 10^{57}$\end{tabular} \\
        ORC6 & 
        \begin{tabular}[c]{@{}l@{}}155.7 $\pm$ 12.8 (M2)\\ 140.8 $\pm$ 8.7 (M4)\end{tabular} &
        \begin{tabular}[c]{@{}l@{}}(1.29 $\pm$ 0.29) $\times 10^{-14}$\\ (1.29 $\pm$ 0.29) $\times 10^{-14}$\end{tabular} &
        \begin{tabular}[c]{@{}l@{}}(8.02 $\pm$ 3.45) $\times 10^{-7}$\\ (2.02 $\pm$ 0.87) $\times 10^{-6}$\end{tabular} &
        \begin{tabular}[c]{@{}l@{}}202.5 $\pm$ 7.5\\ 157.5 $\pm$ 7.5\end{tabular} &
        \begin{tabular}[c]{@{}l@{}}(2.57 $\pm$ 0.98) $\times 10^{58}$\\ (2.57 $\pm$ 0.98) $\times 10^{58}$\end{tabular} \\
    \enddata
\end{deluxetable*}
The resulting distributions suggest:\\
(1) Age constraint: 

From the requirement that the simulated ring reproduces the observed aspect ratio, acceptable solutions exist only for rings younger than $35$ shock-crossing times for Mach 4 shocks or $20$ shock-crossing times for Mach 2 shocks. However, an additional and more restrictive constraint arises from matching the observed radio emissivity. Using the inferred environmental conditions, this constraint limits the ORC ages to $70$–$200$ Myr, corresponding to $\lesssim 5 t_{\rm cross}$, with the broader $1\sigma$ likelihood range allowing ages $25$–$215$ Myr. Although the hydrodynamic ring-like structure can persist for many shock-crossing times, only rings between these limits simultaneously satisfy both the morphological and spectral flux constraints and are therefore consistent with the observed ORCs.
\\
(2) Total energy: The most-likely total initial energies needed to inflate the bubble to the initial bubble size to give the observed ring properties are in the range $10^{58}$–$10^{59}$ erg. Including the broader likelihood intervals, the possible range of total energies extends from $\sim10^{57}$ erg to $\times10^{59}$ erg. These energies are typical of powerful AGN jet outbursts that produce extended radio lobes or cavities in the surrounding medium.\\
(3) Ambient conditions: The most-likely ambient pressures inferred from the fits range within $10^{-14}$~dyn~cm$^{-2}$, while the corresponding number densities are in the range of $10^{-7}$–$10^{-6}$~cm$^{-3}$. When considering the full likelihood distributions, the broader possible ranges span $P \sim10^{-15}$–$10^{-13}$ dyn cm$^{-2}$ and $n \sim10^{-7}$–$10^{-5}$ cm$^{-3}$. These densities are one to two orders of magnitude below the canonical values for the warm–hot intergalactic medium (WHIM; $n \sim 10^{-5}$–$10^{-4}$~cm$^{-3}$), and far below those of intracluster gas ($n \gtrsim 10^{-3}$~cm$^{-3}$). The pressures are comparable to the typical conditions of the diffuse intergalactic medium (IGM; $10^{-15}$–$10^{-14}$~dyn~cm$^{-2}$), and well below those found in the hot intracluster medium (ICM; $P \sim 10^{-12}$–$10^{-10}$~dyn~cm$^{-2}$). 
Such low-density, moderate-pressure conditions suggests that the observed ORCs are likely expanding within environments at the interface between the diffuse intergalactic medium (IGM) and the outermost circumgalactic or intragroup medium.


\subsubsection{Polarization Properties}\label{sec:polarization}
Our simulations of shock-bubble interactions yield a polarization fraction of approximately 20-40\%, notably below the theoretical maximum of 75\% expected for synchrotron emission from a perfectly ordered magnetic field. This predicted range is consistent with the available polarization measurement for ORC1\citep{norris_meerkat_2022}, which is the only ORC with published polarization data. The agreement suggests again that moderate–Mach-number shocks (Mach~2–4) are sufficient to reproduce both the level and morphology of the observed polarization. Several factors reduce this value in practice. The magnetic field within our simulated bubbles is initialized as randomized, reflecting the uncertainty in the seed field geometry during bubble inflation, while shock compression generates turbulence and shear, which further disturbs field ordering and introduces significant in-beam depolarization. 

In \S\ref{sec:method}, we identified a clear correlation between polarization properties and ring morphology: higher polarization fractions occur when the ring aspect ratio is larger, and, similarly, the magnetic field orientation is {\em more tangential} to the ring edge when the ring {\em aspect ratio is larger}. For ORC1, the ring radius is 41 arcsec and the synthesized beam is 11 arcsec. 

By eye, We estimate its aspect ratio to be about 2.4, corresponding to a width of 17 arcsec. This places it in a higher-aspect-ratio regime than ORC5 whose ring radius is 35 arcsec and aspect ratio is about 1.4 with a beam size of 13 arcsec, making the ORC5 width 25 arcsec. In our simulations, a higher aspect ratio corresponds to more tangential magnetic fields and higher intrinsic polarization fractions, consistent with the tangential polarization pattern observed in ORC1. While there's no polarization data for ORC5, we predict that thicker rings like ORC5 correspond to more radial or disordered field configurations and lower polarization fractions, providing a direct diagnostic for future polarimetric observations.


However, both observational and numerical resolution effects must be considered when comparing models to data, which we will discuss in the next section.

The absence of polarization detections in other ORCs likely stems from observational limitations. The intrinsic faintness of ORCs ($S_\nu \lesssim 10$ mJy) leads to low polarized flux densities that often fall below current instrument sensitivities. Additionally, in-beam depolarization caused by large synthesized beam sizes averages over unresolved magnetic field structure, further reducing observed polarization fractions. Our simulations confirm that higher angular resolution reveals more complex, small-scale field structures and generally reduces the measured polarization fraction (see \S\ref{sec:analysis}).

Polarization properties thus emerge as a promising diagnostic for distinguishing between different ORC formation models. The relationship between aspect ratio and polarization angle offers a specific, testable prediction: ORCs should display a more radially ordered magnetic field when the aspect ratio is large. 
Future observations will be crucial for understanding whether ORC1's polarization properties reflect its unique environment or represent a broader class of ORC evolution.

\subsection{Caveats}\label{sec:caveats}
As with any numerical study, out simulations are subject to limitations that are important to be aware of when comparing results to observations, which carry their own set of observational limits and biases.

\subsubsection{Image Resolution}

Observational facilities like ASKAP, with a synthesized beam of arcseconds, cannot resolve fine structural details in ORCs, such as the sharpness of the ring edges, the intrinsic width of the synchrotron-emitting regions, and small-scale magnetic field fluctuations. This blurring effect inherently introduces uncertainties in measuring key morphological properties, such as the ring's radial profile and curvature, which are essential to distinguish shock-driven models from alternative mechanisms (e.g., large scale shocks from starburst winds or cosmic ray driven bubbles.) 

We account for this limitation by convolving all synthetic synchrotron maps with the appropriate telescope beam for each ORC. This approach is functionally equivalent to deconvolving the observed image to determine the intrinsic ring width. Both methods serve to place the models and observations on a common resolution footing and enable a direct comparison.

Different telescopes provide different resolutions which shape the observed polarization fraction, flux density, and magnetic field structure. As demonstrated in Section 3, these quantities are highly sensitive to the chosen resolution, which, in turn, affects the interpretation of the data. A factor-of-two improvement in resolution (e.g., from 13 to 6 arcseconds) could decrease the ring’s measured width, directly impacting physical inferences. For instance, a broader inferred ring width might erroneously suggest a slower shock velocity, or an older ring.

Future studies should follow up observations of ORCs at higher angular resolutions. 
Instruments like the Square Kilometre Array may be able to set better constraints on the aspect ratio, one of the most robust diagnostics we find from the simulations.

\subsubsection{Numerical resolution}

A fundamental limitation inherent to grid-based fluid dynamics simulations, including magnetohydrodynamic (MHD) studies like ours, is numerical diffusivity. Numerical diffusivity refers to the artificial diffusion of physical quantities across interfaces that are supposed to remain sharp in the physical system. This occurs due to the discrete nature of computational grids, which can blur sharp discontinuities or boundaries such as the boundary of the bubble and the shock. We have shown the results in the analysis section where numerical mixing possibly plays a crucial role in the evolution of magnetic field strength and flux density. 

To mitigate this, we employed adaptive mesh refinement (AMR), which dynamically increases grid resolution in regions of interest, forcing the highest resolution inside the bubbles and at their boundaries. Despite this, numerical diffusivity remains a concern, particularly as the bubble evolves. While the initial bubble is well-resolved, the shock compression significantly reduces its volume. Due to flux freezing, the same amount of magnetic flux is confined into a smaller region with more field reversals. Consequently, even with AMR, we have fewer cells to resolve this newly compressed and highly tangled magnetic field, leading to increased numerical dissipation in these critical regions.

The impact of numerical diffusivity is evident in the resolution-dependent and resolution-independent parameters identified in Fig.~\ref{fig:differentres}. Macroscopic properties like ring width and radius remain robust across resolutions. In contrast, the polarization fraction, magnetic field angle and strength, and integrated flux density, quantities sensitive to the small-scale field structure inside the compressed bubble, vary significantly with resolution. Because of this, we specifically devised our synchrotron analysis to mitigate numerical diffusivity by allowing us to vary the value of the plasma beta parameter when generating synthetic synchrotron maps.

In future work, it will be important to explore higher-resolution simulations or more advanced numerical schemes to better capture the magnetic field evolution in these highly compressed regions and minimize numerical dissipation.

\subsubsection{Host galaxy uncertainty and Frequency}
A critical assumption in this study is the reliance on host galaxy redshifts to infer the distances and physical scales of the ORC rings. While the observed rings\footnote{Here, host galaxy refers to the galaxy from which the initial radio lobe/bubble originated.} are spatially associated with central galaxies (e.g., ORC5 and its host at z=0.27), no direct spectroscopic or photometric redshift measurements of the diffuse synchrotron-emitting rings themselves is possible. This introduces systematic uncertainty in the precise distances and thus sizes of the observed rings $R_{\text{ring}}\propto D_A(z)$, as well as $E_\text{tot} \propto D_L(z)^2/(1+z)^{4+\alpha}$ as the true redshift may differ from the presumed host. We can speculate on a scenario where the central galaxy is not the host, which would imply the ORC is either a smaller, less energetic foreground object or a larger, more powerful background event. 
However, our formation model is robust against this specific uncertainty. Its primary requirement is a shock wave viewed nearly face-on, a condition that does not need a connection to the central galaxy. The ORC could instead be an intergalactic shock viewed in projection with a foreground or background galaxy. Thus, while redshift data is needed to measure absolute values of the physical parameters, our model's validity is not strictly tied to knowing the host galaxy's location.


\subsubsection{Spectral Modeling}
Our approach to a scale-free calculation of the synchrotron spectrum comes with its own set of limitations, as it assumes that one of the two radiative loss mechanisms dominates over the other. In cases where both are important at different times of the simulation, our estimated flux densities must be regarded as upper limits. In follow up work, this could be mitigated by matched re-simulations for the most likely set of parameters drawn from the posterior distributions derived from the spectral modeling presented here. In essence, this would be an iterative approach to finding the most likely numerical model to explain the observations. This is beyond the scope of this paper.

In addition, our calculation of the synchrotron emission and of synchrotron cooling is affected by the numerical dissipation of magnetic energy. We cannot easily mitigate these without resorting to re-simulations with much higher resolutions, once again using the most likely set of parameters drawn for a single ORC (e.g., ORC5.) However, the resolution required for the magnetic dissipation to be fully suppressed would exceed our access to computing time by at least an order of magnitude and we will leave such an investigation to future studies.


\subsubsection{Selection Bias}
The detectability of ORC-like objects is subject to selection effects related to physical scale and viewing angle as we have discussed earlier. Matching the observed sizes of ORCs implies fossil radio bubbles with characteristic diameters of a few hundred kiloparsecs. Such large, highly evolved bubbles are expected to be rare, while smaller bubbles likely produce fainter or less distinct features that are more difficult to identify as ORCs, due to angular resolution limits of the wide area surveys that have produced the first set of ORCs, biasing detections toward the largest systems.

In addition, a well-centered circular ring morphology only happens in a particularly subset of viewing configurations. A nearly perfect circular ring requires the shock surface to be viewed close to face-on, and if the central parent active galaxy of the radio bubbles is also required to appear at the geometric center of the ring, the allowed viewing range becomes even narrower because near-perfect alignment along the line of sight is needed. Such systems are therefore expected to be intrinsically rare but observationally easiest.
However, the SBI model explored in this paper does not {\em require} the parent galaxy to lie at the center. Relaxing this requirement broadens the observable viewing angle range.

As discussed above, recognizable ring-like morphologies are expected for inclinations within 56.5$^{\circ}$, corresponding to roughly 45\% of all possible viewing directions.
At larger inclinations, the same physical model produces distorted or asymmetric structures that may resemble arcs, relics, or filamentary emission rather than clean rings. Because most currently identified ORCs exhibit relatively symmetric morphologies, we have focused here on the nearly face-on cases. A systematic exploration of other predicted morphologies will become more important as more ORC samples are available in the future.

Interestingly, ORC1\citep{norris_odd_2021} exhibits a roughly symmetric double-ring morphology, which in the shock–bubble scenario naturally corresponds to a shock interacting with a pair of fossil bubbles associated with a double-lobed radio galaxy. In this case, the appearance of two distinct rings requires a tilted line of sight and different sizes of initial bubbles so that the near-side and far-side bubbles separate in projection. While we only present the detailed results for a single bubble interacting with a shock in this work, the SBI model naturally predicts this type of system.

In this case, the SBI model places an additional physical constraint on the maximum separation between the two bubbles. For both rings to be visible simultaneously, the time interval between the shock encountering the first bubble and the second bubble must be shorter than the observable lifetime of the individual ring. Using the visible ring lifetimes inferred from our simulations (approximately 5 shock crossing times), the largest possible bubble separation is of order a few of bubble radii. For ORC1, taking an initial bubble radius 240kpc and a Mach 4 shock, this gives the max separation $d_{max} \approx 1.2$Mpc (certainly a plausible constraint even for giant radio galaxies.) If the two bubbles are separated by a larger distance, the first ring is expected to fade before the shock reaches the second bubble, preventing the appearance of a simultaneous double-ring morphology.

\subsubsection{Inhomogeneous Conditions}
Because the fossil bubble can extend across a substantial fraction of the Virial radius of the host halo, the surrounding medium can be inhomogeneous, with varying density, pressure, and temperature. In such environments, physically relevant shocks are likely to be curved and spatially inhomogeneous as they propagate through background gradients, rather than planar and uniform as assumed in our idealized setup. These effects will locally modify the shock strength and lead to additional vorticity generation.

As a consequence, SBI-generated ORCs may be expected to exhibit non-uniform thickness, brightness variations, possibly fragmentation into arcs or filaments, and reduced polarization coherence due to line-of-sight mixing. The first two effects are already seen even in our idealized setup. With inhomogeneous environmental conditions, they are expected to show even stronger variations. We note that the impact of environmental conditions is partially explored through the range of ambient density and temperature parameters considered in this work.

Forming the fossil lobes of hundreds of kpc scale likely requires relatively quiescent environments over extended periods. In dynamically active environments, bulk motions and cosmological flows may distort the plasma lobes, preventing the formation of such large, well-formed bubbles. 
Additional complexity may arise from the buoyant rise and deformation of fossil bubbles prior to shock interaction, which can in some cases produce ring-like structures even in the absence of a strong external shock, whose age would be limited by the radiative losses they would suffer. We are investigating this case separately in forthcoming work. Moreover, curved shocks themselves may generate vorticity independently of bubble interactions, however, in the absence of non-thermal plasma, this would not lead to the formation of radio emission. A systematic exploration of these more complex configurations is beyond the scope of the present paper and is deferred to future work.

\section{Conclusion} \label{sec:conclusion}
We performed 3D MHD simulations of shock-bubble interactions to model the newly discovered class of Odd Radio Circles (ORCs) as vortex rings created through the Richtmyer-Meshkov Instability (RMI). We find that the rings go through a dynamical evolution that has distinct characteristics that can be constrained observationally, such as through measurements of aspect ratio and polarization properties of the rings. We find strong correlations between the ring widths, ring radii, polarization fractions, and magnetic field orientations in the early stages ($\lesssim$ 20 shock crossing time) after the interaction.

Our simulations are largely scale-free. We find that core observational diagnostics, such as the values for radius, width, and magnetic field orientation, are robust against resolution and similar across a wide range of shock Mach numbers and initial magnetic topology of the bubble.

One key finding is that our model naturally produces polarization data consistent with the observations of ORC1, providing a clear observational signature to distinguish it from other theories. Specifically, a more tangential magnetic field angle implies the ring has a larger radius-over-width ratio and a larger polarization fraction (but still in the observed range). Joint measurements of polarization data and aspect ratio can provide an observational test to distinguish between jet-driven and shock-driven ORC scenarios.

We find that magnetic shear significantly amplifies the total magnetic field strength in the later stages of the evolution, enhancing the synchrotron emission and leading to significant radial component of the projected magnetic field of the ORCs. 

Based on our results, we find that the observations of ORCs are consistent with being synchrotron radio sources resulting from shock-bubble interactions, as also suggested by \citet{shabala_are_2024}, generated by moderate strength shocks with Mach numbers in the range from 2-4. 

Assuming the rings lie at the redshifts of their central galaxies, we find that the ORCs are consistent with initial bubble radii of 140–250 kpc, most-likely ambient densities of $\sim10^{-7}$–$10^{-6}$ cm$^{-3}$ (up to $10^{-5}$ cm$^{-3}$ in the broader likelihood range), pressures of $\sim10^{-14}$ dyn cm$^{-2}$ ($10^{-15}$–$10^{-13}$ dyn cm$^{-2}$ possible), and ring ages of roughly 70–200 Myr (extending to 25–215 Myr for all possible values).

These conditions suggest that the ORCs reside in extremely low-density, moderate-pressure environments characteristic of the outskirts of galaxy groups and clusters. In such regions, fossil AGN lobes are likely re-energized through interactions with large-scale shocks driven by accretion flows, mergers, or renewed AGN activity, producing the observed ring-like synchrotron structures.

While the overall properties of the known of ORCs align with our model, several challenges still remain when comparing to the observations: (1) the redshift ambiguity imposed by assigning the central galaxy as the host of the ORC, compared to any other nearby galaxies, (2) the limits of both imaging resolution and numerical dissipation due to numerical resolution, (3) the limited observed polarization data.

Beyond the detection of larger samples of ORCs, future progress requires advances in observational capabilities that can better resolve faint diffuse objects. Future sensitive polarization observations will be crucial for testing our model's prediction of polarization angle. Similarly, higher resolution simulations would help overcome the limits of numerical dissipation on our ability to accurately follow the evolution of the magnetic field down to small scales.
\citep{koss:25}

%
%
\bibliographystyle{aasjournalv7} 
\bibliography{Radio_circles} 


\end{document}